\documentclass[journal,10pt]{IEEEtran}

\usepackage[T1]{fontenc}
\usepackage{textcomp}
\usepackage{xcolor}
\usepackage{amssymb,amsmath,amsfonts,amsthm}
\usepackage{mathrsfs}
\usepackage{cite}
\usepackage{array}
\usepackage{tabularx}
\usepackage{color}
\usepackage{subcaption}
\usepackage{mathtools}
\usepackage{tikz,pgf}
\usetikzlibrary{calc,positioning,mindmap,trees,decorations.pathreplacing}
\usepackage{graphicx}
\usepackage{bbm}
\usepackage{bm}
\usepackage[utf8]{inputenc}
\usepackage{algorithm}
\usepackage{algpseudocode}
\usepackage{stfloats}

\newtheorem{theorem}{Theorem}
\newtheorem{assumption}{Assumption}
\newtheorem{definition}{Definition}

\newtheorem{remark}{Remark}

\newcommand{\matr}[1]{\mathbf{#1}}
\newcommand{\vect}[1]{\mathbf{#1}}

\usepackage[colorinlistoftodos,prependcaption,backgroundcolor=black!5!white,bordercolor=red]{todonotes}

\def\BibTeX{{\rm B\kern-.05em{\sc i\kern-.025em b}\kern-.08em
    T\kern-.1667em\lower.7ex\hbox{E}\kern-.125emX}}
\AtBeginDocument{\definecolor{ojcolor}{cmyk}{0.93,0.59,0.15,0.02}}

\begin{document}

\title{Faster Convergence with Less Communication: Broadcast-Based Subgraph Sampling for Decentralized Learning over Wireless Networks}

\author{{Daniel P\'erez Herrera, Zheng Chen, and Erik G. Larsson}
	
	\thanks{The authors are with the Department of Electrical Engineering, Link\"{o}ping University, Link\"{o}ping, SE-58183 Sweden. E-mail: \{daniel.perez.herrera, zheng.chen, erik.g.larsson\}@liu.se. }
	\thanks{This work was supported in part by Zenith, ELLIIT, Knut and Alice Wallenberg Foundation, and Swedish Research Council (VR). A part of this work was presented at the 2024 IEEE International Conference on Communication (ICC) \cite{herrera2023decentralized}.}
}

\maketitle

\begin{abstract}
Decentralized stochastic gradient descent (D-SGD) is a widely adopted optimization algorithm for decentralized training of machine learning models across networked agents.  A crucial part of D-SGD is the consensus-based model averaging, which heavily relies on information exchange and fusion among the nodes. For consensus averaging over wireless networks, due to the broadcast nature of wireless channels, simultaneous transmissions from multiple nodes may cause packet collisions if they share a common receiver. 
Therefore, communication coordination is necessary to determine when and how a node can transmit (or receive) information to (or from) its neighbors.  In this work, we propose $\texttt{BASS}$, a broadcast-based subgraph sampling method designed to accelerate the convergence of D-SGD while considering the actual communication cost per iteration. 
$\texttt{BASS}$ creates a set of mixing matrix candidates that represent sparser subgraphs of the base topology. In each consensus iteration, one mixing matrix is randomly sampled, leading to a specific scheduling decision that activates multiple collision-free subsets of nodes. The sampling occurs in a probabilistic manner, and the elements of the mixing matrices, along with their sampling probabilities, are jointly optimized.
Simulation results demonstrate that $\texttt{BASS}$ achieves faster convergence and requires fewer transmission slots than existing link-based scheduling methods and the full communication scenario.
In conclusion, the inherent broadcasting nature of wireless channels offers intrinsic advantages in accelerating the convergence of decentralized optimization and learning.
\end{abstract}


\section{Introduction}
\IEEEPARstart{C}{ollaborative} machine learning (ML) with decentralized data has gained widespread recognition as a promising solution for preserving data privacy in large-scale ML. In this approach, models are trained directly on individual devices using locally available datasets. Federated learning is a notable example in this category \cite{mcmahan2017communication}. Traditional studies on federated learning often center around a server-based architecture, wherein locally trained models from distributed agents are periodically aggregated and synchronized at a parameter server. However, this master-worker architecture is susceptible to single node failure and generates high communication load between the server and the agents. An alternative solution to address these challenges is serverless decentralized learning. In this approach, each agent performs local training and communicates directly with its locally connected neighbors for in-network model fusion. 

Training ML models to perform a certain task is essentially solving an optimization problem. One commonly used first-order optimization algorithm for decentralized learning is the Decentralized Stochastic Gradient Descent (D-SGD) \cite{swenson2020distributed, lian2017can}. D-SGD comprises two key components: (1) stochastic gradient descent and (2) consensus-based model averaging. Particularly, the consensus averaging mechanism ensures that the agents can approach a state of agreement on a quantity of interest \cite{olfati2007consensus, xiao2004fast, olshevsky2009convergence}, which is the optimizer of the global objective function.
Convergence analysis of D-SGD has been extensively investigated in the literature \cite{wang2021cooperative, koloskova2020unified, jakovetic2018convergence, nedic2009distributed, scaman2018optimal}. A common understanding is that the communication topology plays a critical role in the convergence performance, which is reflected by the spectral gap of the mixing (or weight) matrix \cite{topology-coomunication-tradeoff}. Recent studies have examined the impact of network topology in the convergence bound and unveiled some counter-intuitive remarks \cite{neglia2020decentralized, vogels2022beyond}. Building on the plain D-SGD design, previous works have explored the effects of compression \cite{koloskova2019decentralized, distributed-event-quantized}, link failures \cite{ye2022decentralized, jiang2022decentralized}, communication delays \cite{lee2022decentralized}, asynchronous updating \cite{wu2023asynchronous}, rate adaptation \cite{sato2021rate}, and knowledge distillation \cite{taya2022decentralized}.

Most existing research on communication-efficient D-SGD focuses on reducing the amount of information transmitted over the links by compressing the communicated models \cite{koloskova2019decentralized, lu2020moniqua, tang2018communication, alistarh2017qsgd}. Some recent works have considered tuning the frequency of communication between nodes, by subgraph sampling \cite{wang2022matcha, chiu2023laplacian}, link scheduling \cite{liu2023communication} or event-triggered communication \cite{singh2021squarm}. 
In this work, we align with the approach in \cite{wang2022matcha, chiu2023laplacian} and exploit the fact that \textit{not all links need to be used in every iteration}. 
Although this approach may slow down the per-iteration convergence rate, activating a subset of nodes and links in each iteration can significantly reduce the communication cost, leading to potential improvement in communication efficiency. 

\subsection{Related Works}
The most closely related work to ours is $\texttt{MATCHA}$ \cite{wang2022matcha}. The key idea is to improve the error-versus-wall-clock-time convergence of D-SGD by activating connectivity-critical links more often, given a fixed communication budget (e.g., number of activated links) per iteration.
Along this line, a Laplacian matrix sampling approach is presented in \cite{chiu2023laplacian}, where a set of candidate Laplacian matrices are designed and sampled with probabilities subject to a given cost measure. 
However, these approaches focus primarily on link-based scheduling and do not account for the potential impact of broadcast transmission. 
In wireless networks featuring broadcast transmission, a node can
effectively reach all of its neighbors at the cost of one
transmission, thereby enhancing information dissemination
throughout the network. If multiple nodes transmit
simultaneously without coordination, packet collisions may occur
when they share a common receiver, leading to information loss.
Although multi-agent optimization with broadcast-based
communication has been considered in \cite{nedic2014distributed, nedic2010asynchronous}, these works focus on convergence analysis and do not optimize
communication dynamics.

In this work, our aim is to provide a systematic approach to optimize the communication pattern and mixing matrix weights for D-SGD with broadcast transmission, which has been overlooked in the current literature. 

\subsection{Contributions}
We propose $\texttt{BASS}$, a communication-efficient framework for D-SGD with broadcast transmission, using probabilistic sampling of sparser subgraphs of the base topology under a given communication cost constraint. Our design consists of three main components: 
\begin{enumerate}
    \item Partitioning the entire network into several collision-free subsets;
    \item Combining multiple collision-free subsets into one candidate subgraph under a given communication cost constraint;
    \item Activating these candidate subgraphs randomly with optimized activation probabilities and mixing weights. 
\end{enumerate}

Furthermore, we present another heuristic approach with reduced computational complexity.
To the best of our knowledge, this is the first work to consider broadcast-based graph sparsification for communication-efficient D-SGD.   
Simulation results show that $\texttt{BASS}$ achieves significant performance gain compared to existing link-based scheduling approaches, validating the benefits of exploiting broadcast transmission in the communication design.

\section{Decentralized Optimization and Learning}

\subsection{Notation}
A vector is represented as $\vect{a} \in \mathbb{R}^d$, and a matrix is represented as $\matr{A} \in \mathbb{R}^{d\times d}$. We use either $(\matr{A})_{ij}$ or $A_{ij}$ to denote the element at the $i$-th row and $j$-th column of matrix $\matr{A}$.
$||\vect{a}||$ is the Euclidean norm of the vector $\vect{a}$, and $||\matr{A}||_2$ is the spectral norm (largest singular value) of the matrix $\matr{A}$. $\mathbbm{1}$ and $\vect{0}$ are column vectors of all-ones and all-zeros, respectively. $\text{ diag}(\vect{a})$ denotes a diagonal matrix whose diagonal entries are the elements of the vector $\vect{a}$. $\lambda_i(\matr{A})$ is the $i$-th smallest eigenvalue of the matrix $\matr{A}$, and $\lambda_{\max}(\matr{A})$ is the largest eigenvalue.

\subsection{Network Model and Graph Preliminaries}
We consider a network of $N$ nodes, modeled as an undirected and connected graph $\mathcal{G}=(\mathcal{V},\mathcal{E})$, with $\mathcal{V}=\{1,\ldots,N\}$ being the set of nodes/agents, and $\mathcal{E}\subseteq\{(i,j)\in \mathcal{V}\times\mathcal{V}|i\neq j\}$ being the set of links/edges.
Each node $i$ can only communicate with its set of neighbors, denoted as $\mathcal{N}_i = \{ j|(i,j)\in \mathcal{E}\}$.
The degree of a node $i$ is $d_i=|\mathcal{N}_i|$, which represents the number of its neighbors.

The topology of the graph $\mathcal{G}$ can be represented by its adjacency matrix $\matr{A}\in \mathbb{R}^{N\times N}$. The elements of $\matr{A}$ are $A_{ij} = 1$ if $(i,j)\in \mathcal{E}$ and  $A_{ij} = 0$ otherwise, with $A_{ii} = 0, \forall i$. 
The Laplacian matrix of $\mathcal{G}$ is defined as $\matr{L} = \matr{D} - \matr{A}$, where $\matr{D}=\text{diag}(d_1,...,d_N)$ is a diagonal matrix whose diagonal elements are the node degrees. We will refer to $\matr{D}$ as the degree matrix of the graph $\mathcal{G}$.
The Laplacian matrix has the sum of each row/column equal to zero, i.e., $\matr{L}\mathbbm{1}=\vect{0}$, and $\mathbbm{1}^{\top}\matr{L}=\vect{0}$.

\subsection{Collaborative Machine Learning in Decentralized Setting}
\label{requirements_W}

Let $\vect{x}\in\mathbb{R}^d$ represent the learning model. The objective of collaborative training can be written as the following optimization problem 
\begin{equation}
	\min_{\vect{x}\in\mathbb{R}^d} \left(F(\vect{x}):=\frac{1}{N}\sum_{i=1}^N F_i(\vect{x})\right),
\end{equation}
where each $F_i:\mathbb{R}^d\rightarrow\mathbb{R}$ defines the local objective function of agent $i$. For model training, the local objective $F_i$ can be defined as the empirical local risk/loss function
\begin{equation}
	F_i(\vect{x}):=\frac{1}{|\mathcal{D}_i|}\sum_{s\in\mathcal{D}_i}f(\vect{x};s),
\end{equation}
where $\mathcal{D}_i$ refers to the local data of agent $i$, and $f(\vect{x};s)$ is the loss function of the learning model $\vect{x}$ for sample $s$.

To perform collaborative model training in decentralized systems, we use decentralized SGD (or D-SGD), that relies on local gradient updating and consensus-based model averaging. Every iteration of the algorithm includes the following steps:
\begin{enumerate}
	\item \textbf{Stochastic gradient update:} Each node $i$ computes its stochastic gradient vector $\vect{g}_i(t)$ and updates its local model according to 
	\begin{equation}
		\vect{x}_i\left(t+\frac{1}{2}\right) = \vect{x}_i(t)-\eta \vect{g}_i(t),
	\end{equation} 
	where $\eta$ denotes the learning rate and $t$ is the iteration index. The stochastic gradient of each node is computed based on a mini-batch of samples drawn from the local dataset, i.e., $\vect{g}_i(t)=\frac{1}{|\xi_i|}\sum_{s\in\xi_i}\nabla f(\vect{x}_i;s)$ with $\xi_i\subseteq \mathcal{D}_i$.
	\item \textbf{Consensus averaging:} Each node shares its local model with its neighbors and obtains an averaged model as
	\begin{equation}
		\vect{x}_i(t+1)=\sum_{j=1}^N W_{ij}(t)\vect{x}_j\left(t+\frac{1}{2}\right),
	\end{equation}  
	where $W_{ij}(t)$ is the weight that node $i$ assigns to the model update received from node $j$. 
\end{enumerate}

We can put the weights in a matrix $\matr{W}(t)\in \mathbb{R}^{N\times N}$ where the $(i,j)$-th element is  $W_{ij}(t)$. This matrix is often referred to as the mixing matrix, or the consensus matrix. 
Note that the mixing matrix can vary through iterations as a result of time-varying topology or communication scheduling.

For D-SGD with time-varying random topology, common assumptions in the literature to guarantee convergence  are the following \cite{wang2022matcha, chiu2023laplacian, ye2022decentralized}:

\begin{assumption}
	Each local objective function $F_i(\vect{x})$ is differentiable and its gradient is $l$-Lipschitz, i.e., $\left\|\nabla F_i(\vect{x})-\nabla F_i(\vect{y})\right\|\leq l\left\|\vect{x}-\vect{y}\right\|, \forall i \in \{1,2,...,N\}$.
\end{assumption}

\begin{assumption}
	Stochastic gradients at each node are unbiased estimates of the true local gradient of the local objectives, i.e., $\mathbb{E}[\vect{g}_i]=\nabla F_i(\vect{x}_i)$.
\end{assumption}

\begin{assumption}
	The variance of the stochastic gradient at each node is uniformly bounded, i.e., $\mathbb{E}[\left\|\nabla f(\vect{x},s)-\nabla F_i(\vect{x})\right\|^2]\leq \sigma^2, \forall i\in \{1,2,...,N\}$.
\end{assumption}	

\begin{assumption}
	The deviation between the local gradients and the global gradient is bounded by a non-negative constant, i.e., $\frac{1}{N}\sum_{i=1}^{N}\left\|\nabla F_i(x)-\nabla F(x)\right\|^2\leq \zeta^2$.
\end{assumption}	

\begin{assumption}
	The mixing matrices $\{\matr{W}(t)\}_{t\in[T]}$ are independently and identically distributed over time, and satisfy $\left\|\mathbb{E}[\matr{W}^{\top}(t)\matr{W}(t)] - \matr{J}\right\|_2<1$, where $\matr{J}=\frac{1}{N}\mathbbm{1}\mathbbm{1}^{\top}$.
	Every mixing matrix $\matr{W}(t)$ is symmetric with each row/column summing to one, i.e., $\matr{W}^{\top}(t) = \matr{W}(t), \matr{W}(t)\mathbbm{1} = \mathbbm{1}, \forall t$.
\end{assumption}	
Note that in Assumption 5, the expectation is taken over all sources of randomness in the mixing matrix. This randomness may be caused by uncontrollable link failures or by node scheduling with a probabilistic policy, as considered in this work.
Such a policy makes the mixing matrix time-varying with an independent and identical distribution. In contrast, if the mixing matrix is fixed through the iterations, the spectral norm condition simplifies to $||\matr{W}-\matr{J}||_2<1$. 

The design of the mixing matrix is highly relevant for the convergence speed of D-SGD as stated in the following theorem.

\begin{theorem}
	\label{theorem1}
	\textnormal{\cite[Theorem 2]{wang2022matcha}} Under assumptions $(1-5)$, if the learning rate satisfies $\eta l\leq \min\{1, (\sqrt{\rho^{-1}}-1)/4\}$, where $\rho:= \left\|\mathbb{E}[\matr{W}^{\top}(t)\matr{W}(t)]-\matr{J}\right\|_2$, then after $T$ iterations:
	\begin{align}
		\frac{1}{T}&\sum_{t=1}^{T}\mathbb{E}\left[\left\|\nabla F(\bar{\vect{x}}(t))\right\|^2\right] \leq  \left[\frac{2(F(\bar{\vect{x}}(1))-F_{\text{inf}})}{\eta T} + \frac{\eta l \sigma^2}{N}\right. \nonumber\\ +
		&\left.\frac{2\eta^2l^2\rho}{1-\sqrt{\rho}}\left(\frac{\sigma^2}{1+\sqrt{\rho}}+\frac{3\zeta^2}{1-\sqrt{\rho}}\right)\right]\frac{1}{1-2D},
	\end{align}
	where $\bar{\vect{x}}(t)=\frac{1}{N}\sum_{i=1}^{N}\vect{x}_i(t)$, $\bar{\vect{x}}(1)$ is the average of the initial parameter vector, $F_{\text{inf}}$ is a lower bound on $F(\cdot)$, and $D=6\eta^2l^2\rho/(1-\sqrt{\rho})^2<1/2$.
\end{theorem}

As a consequence of the previous theorem, the authors of \cite{chiu2023laplacian} showed that the number of iterations required to achieve a certain level of convergence, i.e., the average expected gradient norm smaller than a threshold, is affected by $\rho$, where a smaller $\rho$ means fewer iterations.
Then, for achieving fast convergence, one may design the distribution of the mixing matrices $\{\matr{W}(t)\}_{t\in[T]}$ such that the spectral norm $\rho$ is as small as possible.

\subsection{Discussions on the Properties of the Mixing Matrix}
In the literature, it is common to consider doubly stochastic mixing matrices \cite{ye2022decentralized, xin2020general, xing2021federated}, which implies that all its entries are non-negative, and the sum of every row/column is equal to one. This assumption simplifies the convergence conditions since we can use the results from Perron-Frobenius theorem for primitive matrices. One example of a doubly stochastic mixing matrix design is to choose $\matr{W}=\matr{I}-\epsilon\matr{L}$, with $\epsilon < \frac{1}{d_{\max}}$, where $d_{\max}$ is the maximum degree in the network. However, as stated in \cite{chiu2023laplacian}, the non-negativity of the mixing matrix is not a necessary condition. 

Another commonly imposed condition on the mixing matrix is the symmetry, i.e, $\matr{W}^{\top}=\matr{W}$. With a symmetric matrix, the spectral norm has the relation $\left\|\matr{W}^{\top}\matr{W} - \matr{J}\right\|_2 = \left\|\matr{W} - \matr{J}\right\|_2^2$, which simplifies the optimization problem. 
However, the convergence proof of D-SGD proposed in \cite[Theorem 2]{wang2022matcha} does not require the symmetry condition of the mixing matrices.
In this work, we assume symmetric mixing matrices, but do not require them to be non-negative.

\section{D-SGD Over Wireless Networks With Broadcast Communication}

When implementing D-SGD over wireless networks, we can exploit the broadcast nature of wireless channels. At the cost of one transmission, a node can share its current local model to all its neighbors. However, if all nodes broadcast at the same time, the information reception will fail due to packet collisions. In this work, we focus on the Medium Access Control (MAC)
layer of the communication network, where the key question is
when and how a node can access the channel and transmit information to its neighbors in a given topology.\footnote{For the moment, we do not consider the effects of PHY channel dynamics such as fading and noise. The joint MAC and PHY layer communication design will be investigated in future work.} In general, there are two types of approaches for multiple access communication: 
\begin{itemize}
	\item random access, i.e., nodes make random decisions to access the channel and transmit their information. This policy is easy to implement and requires no centralized coordination, but it is prone to collisions \cite{chen2023decentralized};
	\item perfectly scheduled, i.e., nodes are divided into collision-free subsets and different transmission slots are allocated to different sets to avoid collision. 
\end{itemize}

In this work, we consider the collision-free approach. For any pair of connected nodes $(i,j)$, the information transmission from node $i$ to $j$ is successful (there is no collision) only if all other neighbors of $j$ (and $j$ itself) are not transmitting. This condition imposes constraints on the set of  links that can be simultaneously active.
We define the notion of collision-free subsets as follows:
	\begin{definition}
		A collision-free subset $\mathcal{S}$ is a set of nodes that can
		transmit simultaneously without collisions, i.e., it satisfies
		$\mathcal{N}_i\cap\mathcal{N}_j = \emptyset,
		(i,j)\notin\mathcal{E}, \forall i,j\in\mathcal{S}$.
\end{definition}

\subsection{Iterations vs. Transmission Slots}
In this work, we exploit the benefits of partial communication, where not all nodes/links are activated in every iteration of the D-SGD algorithm. This allows us to reduce the communication cost and improve the convergence speed measured by objective improvement per transmission slot. A similar approach is adopted in \cite{wang2022matcha} and \cite{chiu2023laplacian}.

\begin{figure}[t!]
	\centering
	\includegraphics[scale=0.35]{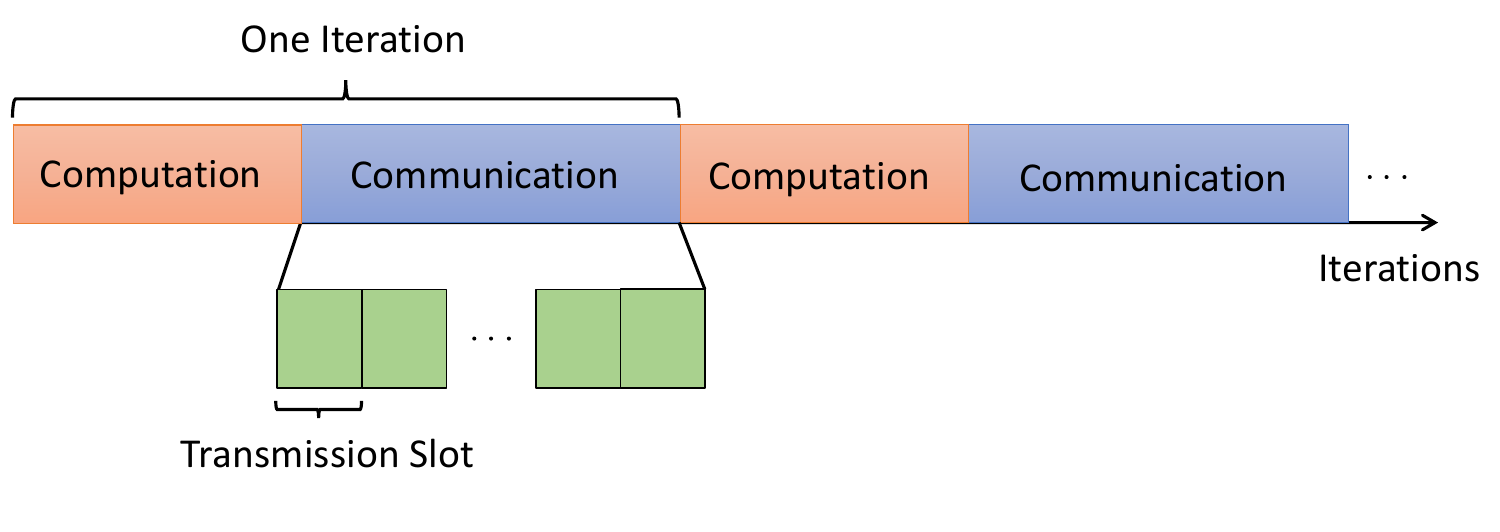}
	\caption{Timeline of the training process. Each iteration consists of one computation phase and one communication round. Multiple transmission slots might be consumed in each communication round, depending on the number of scheduled nodes/links.
	}
	\label{transmission_slots}
\end{figure}

According to our design, in every iteration, one communication round contains multiple transmission slots, as illustrated in Fig. \ref{transmission_slots}. Within each transmission slot, a collision-free subset is activated where all nodes in the subset can broadcast model updates to their neighbors simultaneously. Consequently, packet collisions can be perfectly avoided, but more activated subsets will increase the communication cost/delay per iteration. Here, our measure of time is the number of slots instead of the number of iterations.

\subsection{Link vs. Node Scheduling}
Both \cite{wang2022matcha} and \cite{chiu2023laplacian} consider link-based scheduling.
In our work, the communication cost is defined as the number of transmission slots. With link-based scheduling, one successful information exchange over a bi-directional link requires two transmission slots. In an extreme case without simultaneous transmissions, the communication cost in every iteration will be twice the number of scheduled links in the network. 	
On the other hand, with broadcast transmission, the communication cost will be equal to the number of scheduled broadcasting nodes.
With node-based scheduling and broadcast transmission, we can activate more communication links with fewer transmission slots as compared to link scheduling. 

Considering the aforementioned aspects, we present our proposed communication framework for D-SGD over wireless networks,  namely \texttt{BASS}, using broadcast-based node scheduling and subgraph sampling.

\section{$ \texttt{BASS}$: Broadcast-Based Subgragh Sampling for Communication-Efficient D-SGD}
\label{BASS ext.}
The main idea behind $\texttt{BASS}$ is that the communication cost/delay per iteration can be reduced by selecting a subgraph of the base topology to perform consensus averaging. 
From the mixing matrix perspective, in every iteration, the activated subgraph induces a specific sparsity pattern in $\matr{W}(t)$, i.e., which elements in $\matr{W}(t)$ are non-zero. Then two questions naturally arise in our subgraph sampling design:
\begin{enumerate}
	\item How to determine the structure (sparsity pattern) of the mixing matrix $\matr{W}(t)$?
	\item How to optimize the elements (weights) in $\matr{W}(t)$ given the structure constraint?
\end{enumerate}

Two possible approaches can be developed. The first one is that, in
every iteration $t$, we make the scheduling decision
$\psi(t)$ which decides the set of broadcasting nodes, following a probabilistic policy.\footnote{Theoretically, deterministic policies are also applicable. In this work, we only consider probabilistic policies since the convergence analysis in Theorem 1 requires the mixing matrices to be independently and identically distributed.} Given the scheduling decision $\psi(t)$ and its induced subgraph $\mathcal{G}(t)$, we optimize the mixing matrix $\matr{W}(t)$. The second approach is, we create a family of mixing matrix candidates $\{\matr{W}_r\}_{r\in[R]}$, where each matrix $\matr{W}_r$ corresponds to one subgraph $\mathcal{G}_r$ with a certain communication cost. Then we can design a policy to randomly select the mixing matrix in each iteration $\matr{W}(t)\in\{\matr{W}_r\}_{r\in[R]}$, and activate the corresponding nodes in the network. 

We start with the second approach, as illustrated in Fig.~\ref{Link_removal}. First, nodes are divided into multiple disjoint collision-free subsets. Then, a family of candidate subgraphs are created as the union of multiple activated collision-free subsets under a given communication cost constraint. These subgraphs are randomly sampled in every iteration, and the sampling probabilities and the mixing matrices associated with these subgraphs are jointly optimized.

\subsection{Step 1: Create Collision-Free Subsets}
\label{step1}

In communication networks, information transmission and reception between a pair of nodes is always directed. To reflect the direction of communication, we view the base topology $\mathcal{G}$ as a directed graph $\mathcal{G}^d=(\mathcal{V},\mathcal{E}^d)$ with bi-directional links between each pair of connected nodes. 

With broadcast transmission, each broadcasting node creates a local ``star'' with directed links towards its neighbors. Considering the collision-free conditions, we divide the base topology into disjoint subsets of nodes $\{\mathcal{V}^k\}_{k\in[q]}$, where $q\leq N$ is the total number of collision-free subsets. The elements in $\mathcal{V}^k$ are the nodes that can transmit simultaneously without collision, i.e. $i,j\in\mathcal{V}^k$ if $ \mathcal{N}_i\cap\mathcal{N}_j = \emptyset$, and $(i,j)\notin\mathcal{E}$. 
This partition satisfies $\cup_{k=1}^q\mathcal{V}^k = \mathcal{V}, \mathcal{V}^k\cap\mathcal{V}^l=\emptyset, \forall k\neq l$. 

\begin{figure}[t!]
	\centering
	\includegraphics[width=\columnwidth]{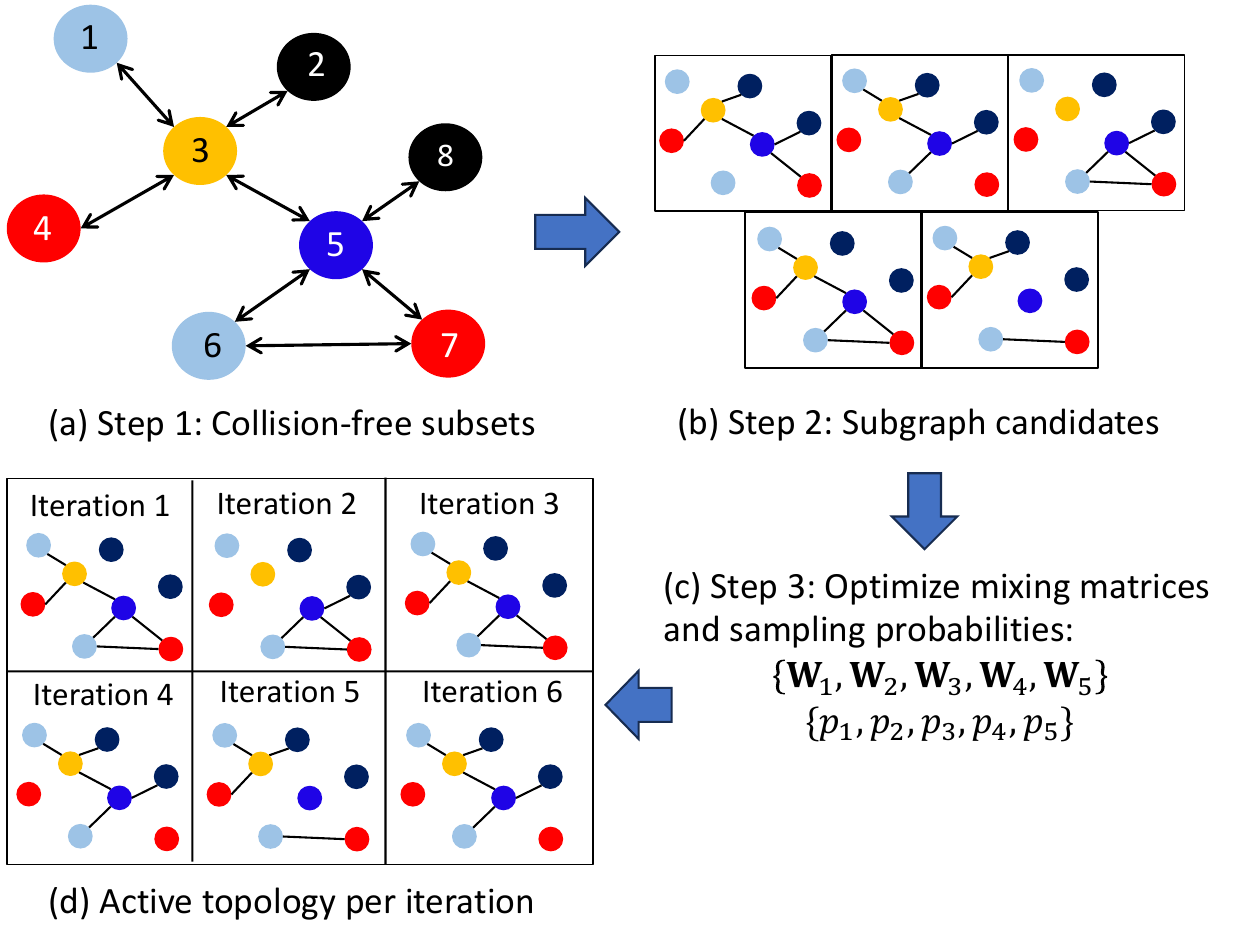}
	\caption{(a) Partition of the base graph into collision-free subsets, where different colors represent different collision-free subsets (b) Subgraph candidates for a communication budget of $\mathcal{B}=4$. (c) Optimization of the mixing matrix candidates and their sampling probabilities. (d) Example of sampled subgraphs per iteration.}
	\label{Link_removal}
\end{figure}

Finding an optimal partition (in the sense of finding a partition with the smallest number of subsets $q$) is a combinatorial optimization problem that, in general, is NP-hard. 
Some heuristic algorithms can be used, such as the greedy \textit{vertex-coloring} algorithm \cite{bollobas1998modern}, that assign different colors to connected nodes.
To address the constraint that two nodes cannot transmit simultaneously (be in the same subset) if they share a common neighbor, we follow a similar approach as in \cite{xing2021federated}. We apply the coloring algorithm over an auxiliary graph $\mathcal{G}^a=(\mathcal{V},\mathcal{E}^a)$ where $\mathcal{E}^a$ includes all edges in $\mathcal{E}$ and an additional edge between each pair of nodes sharing at least one common neighbor.
Then, the nodes marked with the same color will be assigned to the same collision-free subset. The number of subsets $q$ is equal to the number of colors found. 
For the base topology in Fig. \ref{Link_removal}, an example of dividing the base topology into multiple collision-free subsets, where each subset is marked with a different color, is given in Fig. \ref{Link_removal}(a).
The greedy coloring algorithm can be implemented with computational time proportional to the size of the graph $\mathcal{G}^a$, i.e.,  $O(|\mathcal{E}^a|)$ \cite{kubale2004graph}.

\begin{remark}
	The graph partition obtained with the vertex-coloring algorithm is not unique. In this work, our focus is how to schedule the collision-free subsets for a given graph partition. Optimizing the graph partition can be investigated as another independent problem.
\end{remark}

\subsection{Step 2: Create Subgraph Candidates}
\label{step2}

With the collision-free protocol, the number of transmission slots in one iteration is equal to the number of scheduled collision-free subsets.	
For a given communication budget $\mathcal{B}$ (number of transmission slots per iteration), we can find all possible combinations of $\mathcal{B}$ collision-free subsets, obtaining a set $\mathcal{D}$ with $R =$ $q\choose \mathcal{B}$ elements, and each element $D_r, r=1,...,R$, is a collection of $\mathcal{B}$ collision-free subsets. 
For example, if $\mathcal{B}=4$, using the collision-free subsets in Fig. \ref{Link_removal}(a), one element of $\mathcal{D}$ is $D_r = \{ \{2,8\},\{3\}, \{4,7\}, \{5\} \}$, represented by the first subgraph candidate in Fig. \ref{Link_removal}(b).

We further impose that when a node is not scheduled for transmission, it cannot receive information neither. From the consensus averaging perspective, it means that a node will pull an update from a neighbor only when it has pushed an update to that neighbor. 
Then for each $D_r\in \mathcal{D}$, the corresponding subgraph $\mathcal{G}_r=\{\mathcal{V}_r, \mathcal{E}_r\}$  has node set $\mathcal{V}_r$ containing all the nodes of the collision-free subsets in $D_r$, and has link set $\mathcal{E}_r = \{(i,j)\in \mathcal{E}|i,j\in \mathcal{V}_r\}$, i.e.,
all the links associated to the scheduled nodes. 
Note that preserving only the links in the base topology that connect the nodes in $\mathcal{V}_r$ is equivalent to taking the base topology and disconnecting any node $i\notin V_r$. 
However, any non-scheduled node $i$ in iteration $t$ still have access to its own value, therefore, $[\matr{W}(t)]_{ii}=1$.

\begin{remark}
In our design, we impose the constraint that a node cannot receive information when it is not transmitting to maintain the symmetry of the effective communication topology. Therefore, all candidate subgraphs are undirected and all candidate mixing matrices are symmetric. Implementing D-SGD over time-varying directed graphs usually requires a push-sum type mechanism to correct the bias caused by the imbalanced information flow \cite{nedic2009distributed}. This mechanism introduces extra communication cost, as an additional consensus step is performed in every iteration, which doubles the amount of exchanged information. 
\end{remark}

We define the adjacency matrix of the \textit{effective communication topology} as

\begin{equation}
	\matr{A}_r=\matr{Q}_r\matr{A}\matr{Q}_r,
	\label{Atilde}
\end{equation}
where $\matr{Q}_r = \text{diag}(n_1,...,n_N)$, with $n_i=1$ if $i\in\mathcal{V}_r$ and $n_i=0$ otherwise. 
For each $D_r$, we can create an effective communication topology and the adjacency matrix of the subgraph candidate.
An example is given in Fig. \ref{Link_removal}(a)-(b).

\subsection{Step 3: Optimize Mixing Matrices and Sampling Probabilities}
\label{section4c}
In every iteration, a subgraph candidate $\mathcal{G}_r$ is randomly sampled with probability $p_r$, which maps to one particular combination of collision-free subsets $D_r\in\mathcal{D}$. The corresponding scheduling decision is $\psi(t)=D_r$. The sampled subgraph in slot $t$ determines the sparsity pattern of the mixing matrix $\matr{W}(t)$, i.e., which elements are non-zero.

As stated in Theorem 1 in Section \ref{requirements_W}, the spectral norm $\left\|\mathbb{E}[\matr{W}^{\top}(t)\matr{W}(t)]-\matr{J}\right\|_2$ plays a key role in the convergence speed of D-SGD. Therefore, the elements (link weights) inside each mixing matrix candidate should be determined such that $\left\|\mathbb{E}[\matr{W}^{\top}(t)\matr{W}(t)]-\matr{J}\right\|_2$ is minimized. 
The mixing matrix candidates and their activation probabilities can be optimized by solving:
\begin{subequations}
	\begin{align}
		\min_{\matr{W}_1,...,\matr{W}_R, p_1,...,p_R} \quad & \left\|\sum_{r=1}^{R}p_r\matr{W}_r^{\top}\matr{W}_r-\matr{J}\right\|_2 \\
		\textrm{s.t.} \quad & \sum_{r=1}^{R} p_r = 1, \label{c1}\\
        & \sum_{r=1}^{R} |D_r|p_r \leq \mathcal{B}, \label{dynamic_bass}\\
		& 0\leq p_r \leq 1, \forall r, \label{c2}\\
		& \matr{W}_r^{\top}=\matr{W}_r,\forall r, \label{c3}\\
		& \matr{W}_r\mathbbm{1}=\mathbbm{1}, \forall r, \label{c4}\\
		& \matr{W}_r\in \mathcal{M}_r, \forall r. \label{c5}
	\end{align}
	\label{joint_optimization}
\end{subequations}
For the constraints in this problem:
\begin{itemize}
	\item  \eqref{c1} ensures that the sampling probabilities of the mixing matrix candidates sum to one. 
    \item \eqref{dynamic_bass} ensures that, on average, the number of transmission slots used per iteration is equal to $\mathcal{B}$.
	\item \eqref{c2} imposes that the value of the probabilities should be between $0$ and $1$. 
	\item \eqref{c3} and \eqref{c4} ensure that the mixing matrix candidates are symmetric and the rows/columns sum to one, according to Assumption 5 in Sec. \ref{requirements_W}.
	\item \eqref{c5} ensures that each mixing matrix $\matr{W}_r$ is consistent with its corresponding subgraph $\mathcal{G}_r$. This means that any matrix $\matr{W}\in\mathcal{M}_r$ has $W_{ij}>0$ only if $(i,j)\in\mathcal{V}_r$. 
\end{itemize} 

For a given set of mixing matrix candidates, this problem is convex with respect to the probabilities $\{p_r\}_{r\in[R]}$, and it can be formulated as a semidefinite programming (SDP) problem. It is also convex with respect to all $\{\matr{W}_r\}_{r\in[R]}$ for a given set of activation probabilities. 
The proof is given in the appendix.
However, it is not jointly convex with respect to all $\{\matr{W}_r\}_{r\in[R]}$ and $\{p_r\}_{r\in[R]}$. For this reason, we propose to use alternating optimization to find a local minima.

\subsubsection{Optimize Sampling Probabilities}
For a given set of candidates of mixing matrices $\{\matr{W}_r\}_{r\in[R]}$,
we optimize their activation probabilities by solving:
\begin{subequations}	
	\begin{align}
		\min_{s,p_1,...,p_R} \quad & s \\
		\textrm{s.t.} \quad & \sum_{r=1}^{R} p_r \matr{W}_r^{\top} \matr{W}_r - \matr{J} \preccurlyeq s\matr{I},\\
		& \sum_{r=1}^{R} p_r = 1,\\
        & \sum_{r=1}^{R} |D_r|p_r \leq \mathcal{B}, \\
		& 0\leq p_r \leq 1, \forall r.
	\end{align}
	\label{optiming_p_for_candidates_W}
\end{subequations}
\noindent
Note that \eqref{optiming_p_for_candidates_W} is an SDP problem with $R+2$ linear constraints and with matrix dimension equal to $N\times N$, where $N$ is the number of nodes. Therefore, this problem can be solved in time $O(N^{3.5}+\sqrt{N}R^3)$ using interior point methods and standard matrix multiplication \cite{jiang2020faster}.
\subsubsection{Optimize Each Mixing Matrix Candidate} 
\label{section4c2}
For a given set of mixing matrix candidates $\{\matr{W}_r\}_{r\in[R]}$, and their corresponding sampling probabilities $\{p_r\}_{r\in[R]}$, we can optimize each candidate by solving the following problem:
\begin{subequations}
	\begin{align}
		\min_{s,\matr{W}_r} \quad & s \\
		\textrm{s.t.} \quad & \begin{pmatrix}
			s\matr{I}-\matr{Z}_r & \sqrt{p_r}\matr{W}_r \\
			\sqrt{p_r}\matr{W}_r & \matr{I}
		\end{pmatrix} \succcurlyeq 0,
		\\
		& \matr{W}_r^{\top}=\matr{W}_r,\\
		& \matr{W}_r\mathbbm{1}=\mathbbm{1},\\
		& \matr{W}_r\in \mathcal{M}_r \label{9e}.
	\end{align}
	\label{optiming_W_candidate}
\end{subequations}
where $\matr{Z}_r = \sum_{l=1, l\neq r}^{R} p_l\matr{W}_l^{\top} \matr{W}_l - \matr{J}$. We optimize one mixing matrix at a time while fixing the sampling probabilities and the other mixing matrix candidates. This SDP problem has matrix dimension of $2N\times2N$, and \eqref{9e} imposes $N(N-1)/2-|\mathcal{E}_r|$ constraints, where $|\mathcal{E}_r|$ indicates the number of links in each subgraph candidate $\mathcal{G}_r$. Therefore, this problem can be solved in time $O(N^{6.5})$ with interior point methods using standard matrix multiplication \cite{jiang2020faster}. 

The procedure for finding a solution to \eqref{joint_optimization} is summarized in Algorithm \ref{alg:opt_w_cand_and_p}, where $M$ is the number of alternating iterations. 
Algorithm \ref{alg:opt_w_cand_and_p} can be executed at a central entity/coordinator prior to the training process, which then distributes the resulting mixing matrix candidates and their sampling probabilities to the nodes in the network. 
The complexity of Algorithm \ref{alg:opt_w_cand_and_p} is dominated by lines \ref{dom_eq_1} and \ref{dom_eq_2}, executed $MR$ times.
Consequently, its overall complexity is $O\left(MRN^{6.5}+MR^4\sqrt{N}\right)$.

In every iteration of the D-SGD algorithm, all agents use a common source of randomness to obtain a randomly sampled mixing matrix. From the sampled matrix $\matr{W}_r$, the corresponding scheduling decision can be easily determined by checking (for example, in a lookup table also shared by the coordinator) which combination of collision-free subsets has been used to create the corresponding subgraph associated to $\matr{W}_r$. Note that when the base topology changes, the mixing matrix candidates and their sampling probabilities need to be re-computed. 

\begin{algorithm}[t!]
	\caption{Alternating Optimization Algorithm}

\label{alg:opt_w_cand_and_p}

\algrenewcommand\algorithmicrequire{\textbf{Initialize:}}
\algrenewcommand\algorithmicensure{\textbf{Initialize:}}
\algnewcommand\algorithmicforeach{\textbf{for each}}
\algnewcommand\algorithmicdoinparallel{\textbf{do in parallel}}
\algdef{S}[FOR]{ForEachPar}[1]{\algorithmicforeach\ #1\ \algorithmicdoinparallel}
\algdef{S}[FOR]{ForEach}[1]{\algorithmicforeach\ #1\ \algorithmicdo}
\begin{algorithmic}[1]
	\Require 
	\Statex $\{\matr{W}_r\}_{r\in[R]}, \{p_r\}_{r\in[R]}$, $\{\mathcal{M}_r\}_{r\in[R]}, \text{ and } M$.
	\For{$m \text{ in } [1,M]$}
	\For{$r \text{ in } [1,R]$}
	\State Use $\{p_r\}_{r\in[R]}$ to solve (\ref{optiming_W_candidate}) and find $\matr{W}_r^{(m)}$ \label{dom_eq_1}
	\State Update $\{\matr{W}_r\}_{r\in[R]}$ $(\matr{W}_r \leftarrow \matr{W}_r^{(m)})$
	\State Update $\{p_r\}_{r\in[R]}$ solving (\ref{optiming_p_for_candidates_W}) for the new $ \label{dom_eq_2}\{\matr{W}_r\}_{r\in[R]}$
	\EndFor
	\EndFor
	\State \textbf{Return: } $\{\matr{W}_r\}_{r\in[R]}, \{p_r\}_{r\in[R]}$
\end{algorithmic}
\end{algorithm}

\subsection{Choosing Initial Mixing Matrix Candidates}
\label{initial_candidates}
Since (\ref{joint_optimization}) is a non-convex problem, it may be beneficial to use more than one initialization in Algorithm \ref{alg:opt_w_cand_and_p} and choose the best one. 
We consider two possible initialization approaches using the methods proposed in \cite{wang2022matcha} and \cite{chiu2023laplacian}.

\subsubsection{Initialization A}
Suppose we have $R$ subgraph candidates, from which we can obtain the Laplacian of each subgraph as $\matr{L}_r = \matr{A}_r - \matr{D}_r$, where $\matr{D}_r$ is the degree matrix of $\mathcal{G}_r$. 
We define the mixing matrix candidates as:
\begin{equation}
\matr{W}_r=\matr{I}-\epsilon\matr{L}_r,
\label{eps_app}
\end{equation}
where the value of $\epsilon$ needs to satisfy the condition $\left\|\mathbb{E}[\matr{W}^{\top}(t)\matr{W}(t)]-\matr{J}\right\|_2<1$. 
As stated in \cite[Theorem 1]{wang2022matcha}, there exists an $\epsilon$ such that this spectral norm condition is satisfied; therefore, we formulate an optimization problem:	
\begin{subequations}
\begin{align}
	\min_{s, \epsilon, \beta} \quad & s \\
	\textrm{s.t.} \quad & \epsilon^2-\beta\leq 0,\\
	& \matr{I} -2\epsilon\sum_{r=1}^R p_r\matr{L}_r + \beta\sum_{r=1}^R p_r \matr{L}_r^2 -\matr{J}\preccurlyeq s\matr{I}. 
\end{align}
\label{epsopt}
\end{subequations}
\noindent
However, to solve (\ref{epsopt}), we need the sampling probabilities of each subgraph.
As proposed in \cite{wang2022matcha}, we can find the sampling probabilities of the subgraph candidates by maximizing the algebraic connectivity of the expected graph. This can be done by solving the following optimization problem with concave objective function \cite{boyd2006convex}:
\begin{subequations}
\begin{align}
	\max_{p_1,...,p_R} \quad & \lambda_2\left(\sum_{r=1}^{R}p_r\matr{L}_r\right) \\
	\textrm{s.t.} \quad & \sum_{r=1}^{R} p_r = 1,\\
    & \sum_{r=1}^{R} |D_r|p_r \leq \mathcal{B}, \\
	& 0\leq p_r \leq 1, \forall r.
\end{align}
\label{init_prob}
\end{subequations}

Since the base topology is connected, we have that $\lambda_2\left(\matr{L}\right)>0$. 
We also have that $\sum_{r=1}^{R}\matr{L}_r = \mathcal{H}\matr{L}$, where $\mathcal{H}\geq 1$ is a constant. 
If $p_r=\frac{1}{R}$, for $r=1,...,R$, then $\lambda_2\left(\sum_{r=1}^{R}p_r\matr{L}_r\right)=\frac{1}{R}\lambda_2\left(\sum_{r=1}^{R}\matr{L}_r\right)=\frac{\mathcal{H}}{R}\lambda_2\left(\matr{L}\right)>0$. 
This shows that the solution of \eqref{init_prob} will not result in the case where some nodes have zero probability of being scheduled, since in this case we will have $\lambda_2\left(\sum_{r=1}^{R}p_r\matr{L}_r\right)=0$.

Once obtained the sampling probabilities from  \eqref{init_prob}, we can find $\epsilon$ by solving \eqref{epsopt}, and create the set of mixing matrix candidates using (\ref{eps_app}). 
The entire process of $ \texttt{BASS}$ with this initialization approach is summarized in Algorithm \ref{alg: BASS ext.}. 
Note that the outcome of Algorithm \ref{alg:opt_w_cand_and_p} also guarantees that no node will be left out with zero activation probability, since the initialization point already satisfies that $\left\|\mathbb{E}[\matr{W}^{\top}(t)\matr{W}(t)]-\matr{J}\right\|_2<1$, which satisfies the convergence condition of D-SGD.\footnote{When some nodes have zero probability to be activated, the graph is not connected, and we have $\left\|\mathbb{E}[\matr{W}^{\top}(t)\matr{W}(t)]-\matr{J}\right\|_2=1$.} 

\subsubsection{Initialization B}
Inspired by \cite{chiu2023laplacian}, we consider an alternative initialization that might lead to a smaller spectral norm.
One way to represent the Laplacian is $\matr{L} = \matr{B}\matr{B}^{\top}$, where $\matr{B}\in\mathbb{R}^{N\times \Omega}$ is the incidence matrix, and $\Omega = |\mathcal{E}|$ is the total number of edges in the graph. The matrix $\matr{B}$ has elements:
\begin{equation}
\label{incidence_matrix}
B_{ij} = 
\begin{cases}
	1, & \text{if } s(e_j) = i \\
	-1, & \text{if } t(e_j) = i \\
	0, & \text{otherwise}
\end{cases}
\end{equation}	
where $s(e_j)$ and $t(e_j)$ are the starting and ending points respectively of the edge $e_j$ using an arbitrary orientation.
Now, we can define the weighted Laplacian as 
\begin{equation}
\label{weighted_laplacian}
\hat{\matr{L}}=\matr{B} \text{ diag}(\bm{\alpha}) \matr{B}^{\top},
\end{equation}
where $\bm{\alpha}\in\mathbb{R}^{\Omega}$ contains the weights of all edges in the graph.
Using the weighted Laplacian, the mixing matrices are defined as $\matr{W}_r = \matr{I} - \hat{\matr{L}}_r$, where $\hat{\matr{L}}_r$ is the weighted Laplacian of the subgraph $\mathcal{G}_r$.
We can optimize each weighted Laplacian matrix $\hat{\matr{L}}_r$ such that the spectral norm $||\matr{W}_r^{\top}\matr{W}_r - \matr{J}||_2$ is minimized. 
This is equivalent to solving the following optimization problem:
\begin{subequations}
\begin{align}
	\min_{s, \bm{\alpha}_r} \quad & s \\
	\textrm{s.t.} \quad & -s\matr{I} \preccurlyeq \matr{I}-\matr{B}_r\text{ diag}(\bm{\alpha}_r)\matr{B}^{\top}_r - \matr{J} \preccurlyeq s\matr{I},\\
	& \bm{\alpha}_r\geq0.
\end{align}
\label{find_weighted_L}
\end{subequations}	
Note that the candidate subgraphs are not connected due to inactive nodes, thus, the minimum value of the spectral norm $||\matr{W}_r^{\top}\matr{W}_r - \matr{J}||_2$ will be $1$, and the solution to \eqref{find_weighted_L} is not unique. We are only interested in finding one feasible solution.
The difference between this initialization and the previous one is that we do not impose that each mixing matrix has the same off-diagonal elements ($\epsilon$).  
Now that we have a set of mixing matrix candidates, we can solve (\ref{optiming_p_for_candidates_W}) to find their sampling probabilities. This completes the initialization step in Algorithm \ref{alg: BASS ext.}.
Note that lines $1$--$4$ of Algorithm \ref{alg: BASS ext.}, which contain our contribution in this work, are executed only once before the training starts. The complexity is dominated by line 4.

\begin{algorithm}[t!]
\caption{$ \texttt{BASS}$-$\matr{W}$-$p$ training}

\label{alg: BASS ext.}

\algrenewcommand\algorithmicrequire{\textbf{Initialize:}}
\algrenewcommand\algorithmicensure{\textbf{Initialize:}}
\algnewcommand\algorithmicforeach{\textbf{for each}}
\algnewcommand\algorithmicdoinparallel{\textbf{do in parallel}}
\algdef{S}[FOR]{ForEachPar}[1]{\algorithmicforeach\ #1\ \algorithmicdoinparallel}
\algdef{S}[FOR]{ForEach}[1]{\algorithmicforeach\ #1\ \algorithmicdo}
\begin{algorithmic}[1]
	\Require $\mathcal{G}=(\mathcal{V},\mathcal{E})$, $T$ and $\mathcal{B}$.
	\State Find a partition of $\mathcal{G}$ to find the collision-free subsets as described in  Section \ref{BASS ext.}-\ref{step1}
	\State Create the subgraph candidates as in Section \ref{BASS ext.}-\ref{step2}
	\State Create the initialization points as in Section \ref{BASS ext.}-\ref{initial_candidates}
	\State Optimize $\{\matr{W}_r\}_{r\in[R]}$ and $\{p_r\}_{r\in[R]}$ according to Algorithm \ref{alg:opt_w_cand_and_p} \label{dom_eq_alg2}
	\For{$t \text{ in } [0,T]$}
	\ForEachPar{agent/node $i = 1, \dots, N$}
	\State $\vect{x}_i\left(t+\frac{1}{2}\right) = \vect{x}_i(t)-\eta(t) \vect{g}_i(t)$
	\EndFor
	\State Sample a mixing matrix from $\mathcal{W}^*$ to obtain $\matr{W}(t)$
	\State Determine the scheduling decision $\psi(t)$ from $\matr{W}(t)$
	\ForEach{collision-free subset $\mathcal{S}\in\psi(t)$}
	\ForEachPar{agent/node $i\in\mathcal{S}$}
	\State Broadcast model estimate $x_i\left(t+\frac{1}{2}\right)$ 
	\EndFor
	\EndFor
	\ForEachPar{agent/node $i = 1, \dots, N$}
	\State $\vect{x}_i(t+1)=\sum_{j=1}^N {W}_{ij}(t)x_j\left(t+\frac{1}{2}\right)$
	\EndFor
	\EndFor
	
\end{algorithmic}
\end{algorithm}

\subsection{Comparison With Existing Approaches}	
Two closely related works, \cite{wang2022matcha, chiu2023laplacian}, are not directly applicable in the wireless setting, because the partition rule does not satisfy the collision-free condition in wireless networks. 
In addition, even if the partition rule is modified, one link will still require two transmission slots to achieve bi-directional communication. 
In some topologies with densely connected local parts, broadcast-based scheduling can activate many more links as compared to link-based scheduling under the same communication budget. 
Using a partition rule suitable for wireless networks with collision-free communication, optimizing each mixing matrix candidate and their sampling probabilities are the main novel contributions of our work.
Also, we use the methods proposed in \cite{wang2022matcha} to find an initial point for our mixing matrix candidates optimization; therefore, we improve the communication efficiency by further reducing the spectral norm $\left\|\mathbb{E}[\matr{W}^{\top}(t)\matr{W}(t)]-\matr{J}\right\|_2$, leading to faster convergence.

\subsection{Combining \texttt{BASS} With Compressed Updates}
One way to reduce inter-node communication is by tuning the communication frequency of the nodes. With our proposed method $\texttt{BASS}$, we improved the convergence speed of D-SGD over wireless networks by scheduling a sparser subgraph of the base graph in every iteration. 
    Another way to also reduce the inter-node communication is to compress or quantize \cite{chen2024communication, doostmohammadian2024logarithmically} inter-node model updates. Such techniques reduce the amount of data transmitted per iteration, instead of the communication frequency of the nodes.
    Our method could be combined with such complementary techniques to reduce not only the number of transmission slots but also the amount of information transmitted per slot. 

\subsection{Extension to Directed Graphs}
\texttt{BASS} can be implemented for directed graphs with minor modifications in algorithm design. First, each collision-free subset will contain a group of nodes that do not share any common out-neighbor. Second, the symmetry condition $\matr{W}^\top = \matr{W}$  will be removed from the mixing matrix optimization problem. At last, instead of considering D-SGD which is for undirected graphs, other distributed optimization algorithms for directed graphs can be used, such as gradient-push \cite{nedic2014distributed} or DGD-RS \cite{mai2016distributed}.

\section{Simplified Heuristic Design}
The algorithm presented in Section~\ref{BASS ext.} offers a systematic approach to optimize the communication pattern and the mixing matrix design subject to a given communication cost constraint. However, this method entails solving a substantial number of optimization problems, and this number escalates rapidly with increasing network size.
In this section, we propose a simpler heuristic approach to determine the scheduling decisions in every iteration and construct the mixing matrices based on these decisions.

As mentioned earlier, not all nodes are equally important in the network, and more important nodes should be scheduled more often when the communication budget is limited. 
After the graph partition presented in Section \ref{step1}, we have obtained multiple collision-free subsets of nodes. 
Each subset $\mathcal{S}_k$, $\forall k\in[q]$ will be sampled/scheduled with probability $p_{\mathcal{S}_k}$ in every iteration. 
Then we have
\begin{equation}
\mathbb{E}[\matr{Q}(t)]=\text{diag}(p_1,...,p_N),
\end{equation}
where the diagonal elements of $\matr{Q}(t)$ are the scheduling decisions of each node $i\in\mathcal{N}$.
Here, the scheduling probability of a node equals the scheduling probability of the subset to which it belongs, i.e. $p_i = p_{\mathcal{S}_k}$ if $i\in\mathcal{V}_k$.
Since each subset of broadcasting nodes consumes one transmission slot, the number of transmission slots in every iteration is a random number. 
For a given communication budget $\mathcal{B}$, the scheduling probabilities satisfy $\sum_{l=1}^q p_{\mathcal{S}_l} = \mathcal{B}$. 
This quantity indicates the \textit{average} number of transmission slots (average number of scheduled subsets) per iteration. 

Similar to Section \ref{step2}, from a scheduling decision (which collision-free subsets are activated), we create the adjacency matrix of the effective communication topology as $\tilde{\matr{A}}(t)=\matr{Q}(t)\matr{A}\matr{Q}(t)$.
To further simplify the problem, we consider that all links have equal weight $\epsilon$ in the mixing matrix design, meaning that $\matr{W}(t)=\mathbf{I}-\epsilon\tilde{\matr{L}}(t)$\footnote{This way of defining $\matr{W}(t)$ already satisfy the symmetry constraint and that the sum of the rows/columns is equal to one.}, where
\begin{equation}
\label{Lap_epsilon}
\tilde{\matr{L}}(t) = \text{diag}(\tilde{\matr{A}}(t)\mathbbm{1})-\tilde{\matr{A}}(t).
\end{equation} 

The scheduling probabilities can be optimized by solving the following optimization problem:
\begin{subequations}
\begin{align}
	\min_{p_{\mathcal{S}_1},...,p_{\mathcal{S}_q}} \quad & \left\|\mathbb{E}[\matr{W}^{\top}(t)\matr{W}(t)]-\matr{J}\right\|_2 \\
	\textrm{s.t.} \quad & \sum_{l=1}^q p_{\mathcal{S}_l} = \mathcal{B}, \\
	\quad & 0\leq p_{\mathcal{S}_l} \leq 1, \forall l. 
\end{align}
\label{P_colors optimization}
\end{subequations}
However, the expression for $\mathbb{E}[\matr{W}^{\top}(t)\matr{W}(t)]$ includes several nonlinear terms given by products of the probabilities, which makes the problem hard to solve. 
For this reason, we adopt a heuristic approach for designing the scheduling probabilities, based on the idea presented in \cite{herrera2023distributed}. We evaluate the importance of a node in the network by using its betweenness centrality \cite{latora2017complex}. This metric measures the number of shortest paths that pass through each node, and can be computed in time $O(N|\mathcal{E}|)$ \cite{latora2017complex}, where $\mathcal{E}$ is the number of links in the base graph.
Let $b_i$ represent the centrality value of node $i$. Then we define the importance value of a subset as 
\begin{equation}
b_{\mathcal{S}_j} = \sum_{i=1}^{N} b_i \mathbf{1}_{\{i\in\mathcal{V}_j\}},
\end{equation}
where $\mathbf{1}_{\{i\in\mathcal{V}_j\}}$ is an indicator function, which is equal to one if $i\in\mathcal{V}_j$, and zero otherwise.
Finally, we choose
\begin{equation}
\label{betcen_prob}
p_{\mathcal{S}_j} = \min\{1,\gamma b_{\mathcal{S}_j}\}
\end{equation}
where the constant $\gamma$ is chosen such that $\sum_{i=1}^{q}p_{\mathcal{S}_i}=\mathcal{B}$.

Given the scheduling probabilities, the only parameter that can affect the spectral norm $\left\|\mathbb{E}[\matr{W}^{\top}(t)\matr{W}(t)]-\matr{J}\right\|_2$ is the link weight $\epsilon$. Similar to \cite{wang2022matcha}, we find the optimal value of $\epsilon$ by solving the following  problem:
\begin{subequations}
\begin{align}
	\min_{s, \epsilon, \beta} \quad & s \\
	\textrm{s.t.} \quad & \epsilon^2-\beta\leq 0,\\
	& \matr{I} -2\epsilon\mathbb{E}\left[\tilde{\matr{L}}(t)\right] + \beta\left(\mathbb{E}\left[\tilde{\matr{L}}^{\top}(t)\tilde{\matr{L}}(t)\right]\right) -\matr{J}\preccurlyeq s\matr{I}, 
\end{align}
\label{epsilon_opt}
\end{subequations}
which is a convex problem, with $s$ and $\beta$ being auxiliary variables. Problem \eqref{epsilon_opt} can be solved in time $O(N^{3.5})$ using interior point methods \cite{jiang2020faster}.
If we define a function $\phi:\mathbb{R}^2\rightarrow\mathbb{R}$, with $\phi(i,j)=1$ if nodes $i$ and $j$ are in the same subset, and zero otherwise, we can express the $(i,j)$-th element of $\mathbb{E}\left[\tilde{\matr{L}}(t)\right]$ as in (\ref{ELt}). Note that $\max\{\phi(i,j),p_j\}=p_j$ if nodes $i$ and $j$ are in different subsets and $\max\{\phi(i,j),p_j\}=1$ if not. 
Similarly, we can express the $(i,j)$-th element of $\mathbb{E}\left[\tilde{\matr{L}}^{\top}(t)\tilde{\matr{L}}(t)\right]$ as in (\ref{EL2})-(\ref{last_eq}).
Algorithm \ref{alg:heuristic_BASS} summarizes the entire process of this heuristic approach.

\begin{algorithm}[t!]
\caption{$ \texttt{BASS}$-$\epsilon$ training}

\label{alg:heuristic_BASS}

\algrenewcommand\algorithmicrequire{\textbf{Initialize:}}
\algrenewcommand\algorithmicensure{\textbf{Initialize:}}
\algnewcommand\algorithmicforeach{\textbf{for each}}
\algnewcommand\algorithmicdoinparallel{\textbf{do in parallel}}
\algdef{S}[FOR]{ForEachPar}[1]{\algorithmicforeach\ #1\ \algorithmicdoinparallel}
\algdef{S}[FOR]{ForEach}[1]{\algorithmicforeach\ #1\ \algorithmicdo}
\begin{algorithmic}[1]
	\Require $\mathcal{G}=(\mathcal{V},\mathcal{E})$, $T$ and $\mathcal{B}$.
	\State Find a partition of $\mathcal{G}$ to find the collision-free subsets as described in Section \ref{BASS ext.}-\ref{step1}
	\ForEach{subset $\mathcal{S}_i$}
	\State Assign a sampling probability $p_{\mathcal{S}_i}$ as in (\ref{betcen_prob})
	\EndFor
	\State Solve (\ref{epsilon_opt}) to determine $\epsilon$
	\For{$t \text{ in } [0,T]$}
	\ForEachPar{agent/node $i = 1, \dots, N$}
	\State $\vect{x}_i\left(t+\frac{1}{2}\right) = \vect{x}_i(t)-\eta(t) \vect{g}_i(t)$
	\EndFor
	\State Make a scheduling decision $\psi(t)$ (sample subsets)
	\State Create $\tilde{\matr{L}}(t)$ as in (\ref{Lap_epsilon})
	\State Construct $\matr{W}(t)=\mathbf{I}-\epsilon\tilde{\matr{L}}(t)$
	\ForEach{$\mathcal{S}\in\psi(t)$}
	\ForEachPar{agent/node $i\in\mathcal{S}$}
	\State Broadcast model estimate $x_i\left(t+\frac{1}{2}\right)$ 
	\EndFor
	\EndFor
	\ForEachPar{agent/node $i = 1, \dots, N$}
	\State $x_i(t+1)=\sum_{j=1}^N {W}_{ij}(t)x_j\left(t+\frac{1}{2}\right)$
	\EndFor
	\EndFor
\end{algorithmic}
\end{algorithm}

\begin{figure*}[ht!]	
\begin{equation}
	\left(\mathbb{E}[\tilde{\matr{L}}(t)]\right)_{ij} =
	\begin{cases}
		-p_i\max\{\phi(i,j),p_j\} (\matr{A})_{ij}, & \text{if } i \neq j \\
		\sum_{m=1}^{N} p_i\max\{\phi(i,m),p_m\}(\matr{A})_{im}(\matr{A})_{mi}, & \text{if } i = j
	\end{cases}
	\label{ELt}
\end{equation}

\begin{equation}			
	\mathbb{E}\left[\tilde{\matr{L}}^{\top}(t)\tilde{\matr{L}}(t)\right]  = \mathbb{E}\left[\text{diag}^2(\tilde{\matr{A}}\mathbbm{1})\right] - \mathbb{E}\left[\text{diag}(\tilde{\matr{A}}\mathbbm{1})\tilde{\matr{A}}\right] - \mathbb{E}\left[\tilde{\matr{A}}\text{diag}(\tilde{\matr{A}}\mathbbm{1})\right] + \mathbb{E}\left[\tilde{\matr{A}}^2 \right]
	\label{EL2}
\end{equation}	

\begin{equation}
	\left(\mathbb{E}\left[ \text{diag}^2(\tilde{\matr{A}}\mathbbm{1})  \right]\right)_{ij} =
	\begin{cases}
		0, & \text{if } i \neq j \\
		\sum_{m=1}^{N}\sum_{k=1}^{N} p_i\max\{\phi(i,m),p_m\}\max\{\phi(i,k),\phi(m,k),p_k\}
		(\matr{A})_{im}(\matr{A})_{ik}, & \text{if } i = j
	\end{cases}
\end{equation}	

\begin{equation}
	\left(\mathbb{E}\left[ \text{diag}(\tilde{\matr{A}}\mathbbm{1})\tilde{\matr{A}}  \right]\right)_{ij} =
	\begin{cases}
		\sum_{m=0}^{N}p_i\max\{\phi(i,j),p_j\}\max\{\phi(i,m),\phi(j,m),p_m\}
		(\matr{A})_{ij}(\matr{A})_{im}, & \text{if } i \neq j \\
		0, & \text{if } i = j
	\end{cases}	
\end{equation}		

\begin{equation}
	\left(\mathbb{E}\left[ \tilde{\matr{A}}\text{diag}(\tilde{\matr{A}}\mathbbm{1})  \right]\right)_{ij} =
	\begin{cases}
		\sum_{m=0}^{N}p_i\max\{\phi(i,j),p_j\}\max\{\phi(i,m),\phi(j,m),p_m\}
		(\matr{A})_{ij}(\matr{A})_{jm}, & \text{if } i \neq j \\
		0, & \text{if } i = j
	\end{cases}	
\end{equation}	

\begin{equation}
	\left(\mathbb{E}\left[ \tilde{\matr{A}}^2\right]\right)_{ij} =
	\begin{cases}
		\sum_{m=0}^{N}p_i\max\{\phi(i,j),p_j\}\max\{\phi(i,m),\phi(j,m),p_m\}
		(\matr{A})_{im}(\matr{A})_{mj}, & \text{if } i \neq j \\
		\sum_{m=0}^{N}p_i\max\{\phi(i,m),p_m\}
		(\matr{A})_{im}, &  \text{if } i = j
	\end{cases}	
	\label{last_eq}
\end{equation}	
\hrulefill	
\end{figure*}

\subsection{Discussions on Complexity}
The heuristic approach has two main advantages: the computational cost and the communication overhead.
Regarding the computation complexity, this approach relies on the betweenness centrality measure of the nodes to find the sampling probabilities of the subsets, which is obtained once before the training process starts. Note that another centrality metric could be used instead.
Eventually, only one optimization problem \eqref{epsilon_opt} needs to be solved for obtaining the mixing matrix design, and the solution is valid during the entire training process. Note that the overall complexity of heuristic \texttt{BASS} is dominated by the complexity of solving such problem.
On the other hand, the approach presented in Section \ref{BASS ext.} requires to find all possible combinations of $\mathcal{B}$ subsets, and then find a good initialization point by optimizing $\{p_r\}_{r\in[R]}$ and $\epsilon$. Then, to find the final set of mixing matrix candidates and their activation probabilities, $2R$ optimization problems need to be solved in each step of the alternating optimization algorithm. In total, this approach requires solving $2(RM+1)$ optimization problems, needing more computational power than our heuristic method.

A comparison between the complexity of our algorithms with $\texttt{MATCHA}$ \cite{wang2022matcha} and $\texttt{LMS}$ \cite{chiu2023laplacian} is given in Table \ref{complexity}, where we can see that our heuristic design has a similar complexity as $\texttt{MATCHA}$. 
An exact comparison between the complexity of $\texttt{LMS}$ and optimized $\texttt{BASS}$ is difficult, since it depends on the number of candidates each method uses. However, for the numerical examples in this paper, as in \cite{chiu2023laplacian}, $R^\texttt{LMS}=3091$, while $R<150$, making $\texttt{LMS}$ the most complex of all methods.
    \begin{table}[t!]
    \centering
    {\normalsize
    \begin{tabular}{|c|c|}
        \hline
        Method & Complexity \\ \hline
        Heuristic \texttt{BASS} (ours) & $O\left(N^{3.5}\right)$ \\ \hline
        \texttt{MATCHA} \cite{wang2022matcha} & $O\left(N^{3.5}\right)$ \\ \hline
        \texttt{LMS} \cite{chiu2023laplacian} & $O\left(R^\texttt{LMS}|\mathcal{E}|^{3.5}+(R^\texttt{LMS})^3\sqrt{N}\right)$ \\ \hline
        Optimized \texttt{BASS} (ours) & $O\left(MRN^{6.5}+MR^4\sqrt{N}\right)$ \\ \hline
    \end{tabular}
    }
    \caption{Complexity comparison of different methods. $R^{\texttt{LMS}}$ is the number of subgraph candidates found by \texttt{LMS}, which generally differs from our number of candidates $R$.}
    \label{complexity}
\end{table}

\begin{figure*}[ht!]
\begin{subfigure}[c]{0.27\linewidth}
	\includegraphics[width=\columnwidth]{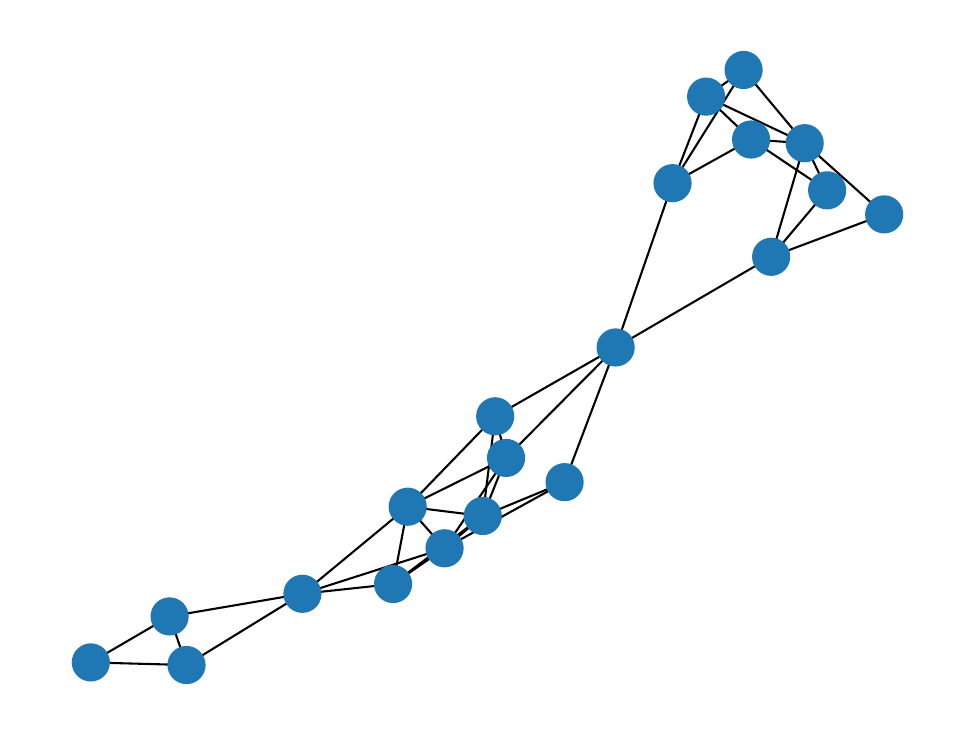}
	\subcaption{Geometric graph}
	\vspace{0.5cm}
\end{subfigure}
\hfill
\begin{subfigure}[c]{0.32\linewidth}
	\includegraphics[width=\linewidth]{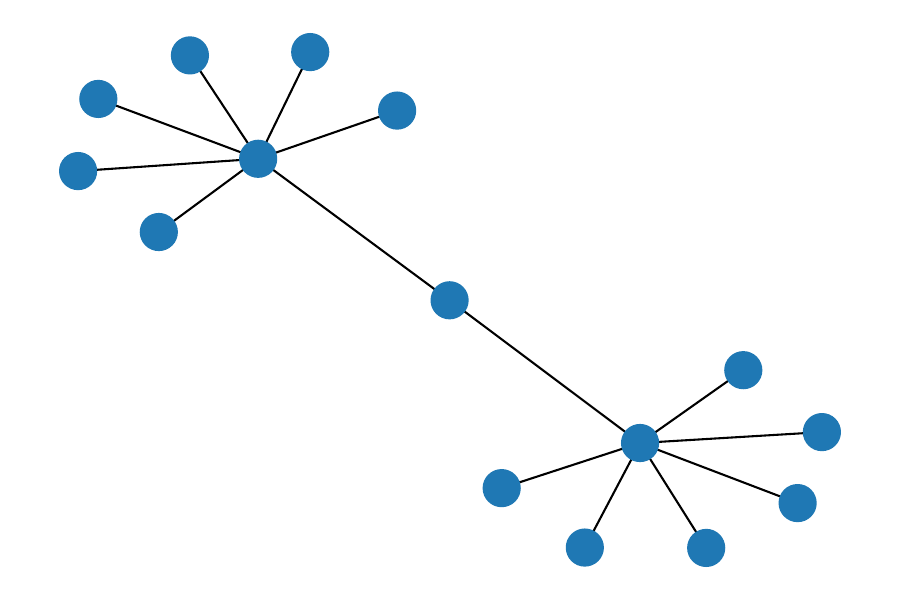}
	\subcaption{Two-stars graph}
	\vspace{0.5cm}
\end{subfigure}		
\hfill
\begin{subfigure}[c]{0.32\linewidth}
	\includegraphics[width=\linewidth]{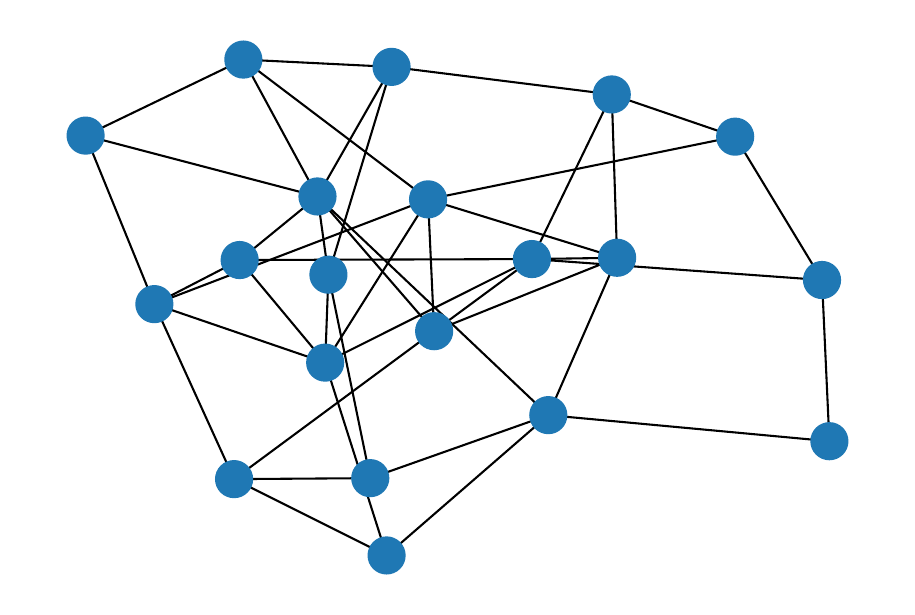}
	\subcaption{Erd\H{o}s-Rényi (ER) graph}
	\vspace{0.5cm}
\end{subfigure}

\begin{subfigure}[c]{0.32\linewidth}
	\includegraphics[width=\linewidth]{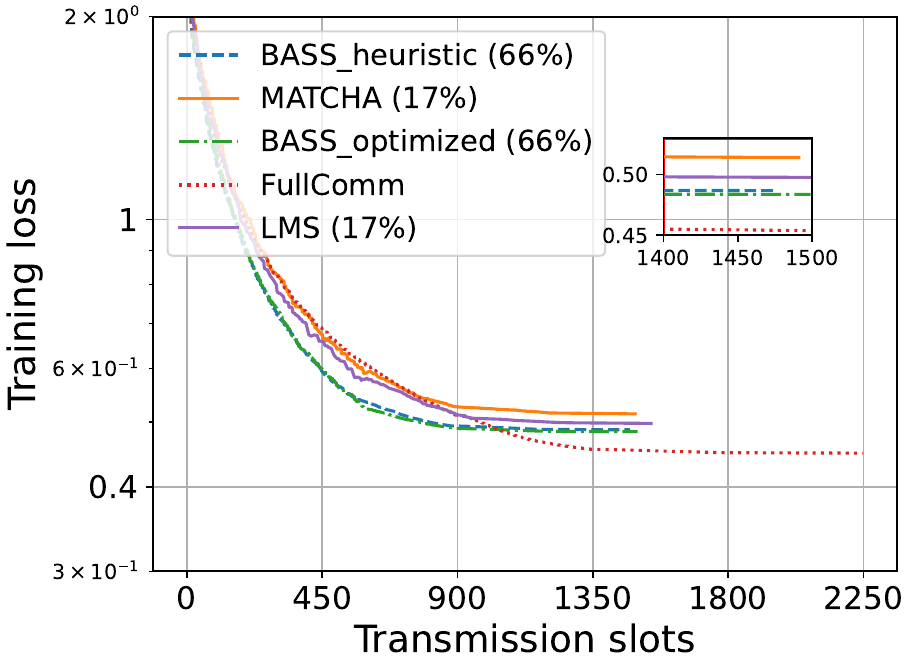}
	\subcaption{Training loss, geometric}
	\vspace{0.5cm}
\end{subfigure}
\hfill
\begin{subfigure}[c]{0.32\linewidth}
	\includegraphics[width=\linewidth]{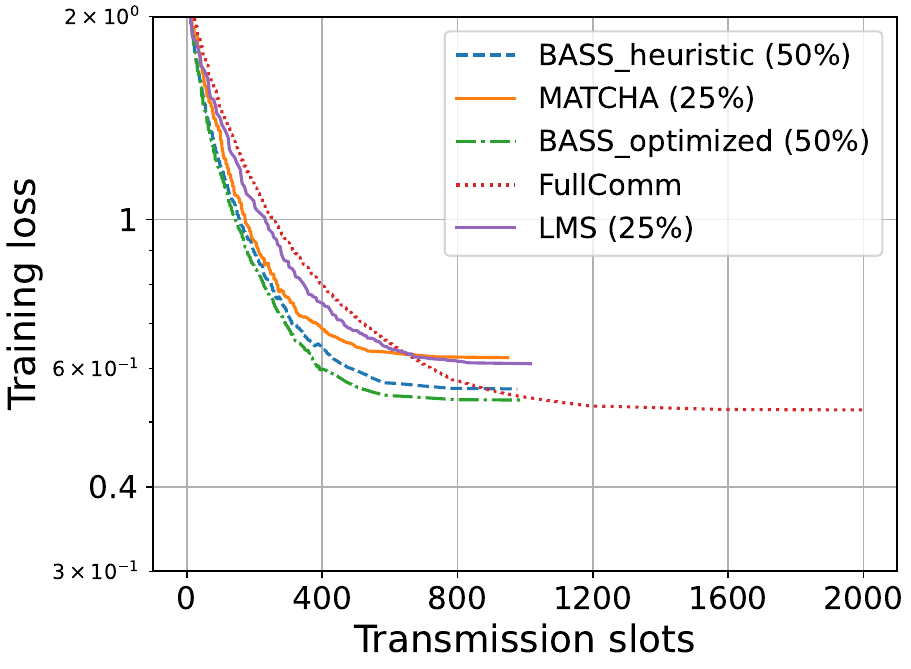}
	\subcaption{Training loss, two-stars}
	\vspace{0.5cm}
\end{subfigure}
\hfill
\begin{subfigure}[c]{0.32\linewidth}
	\includegraphics[width=\linewidth]{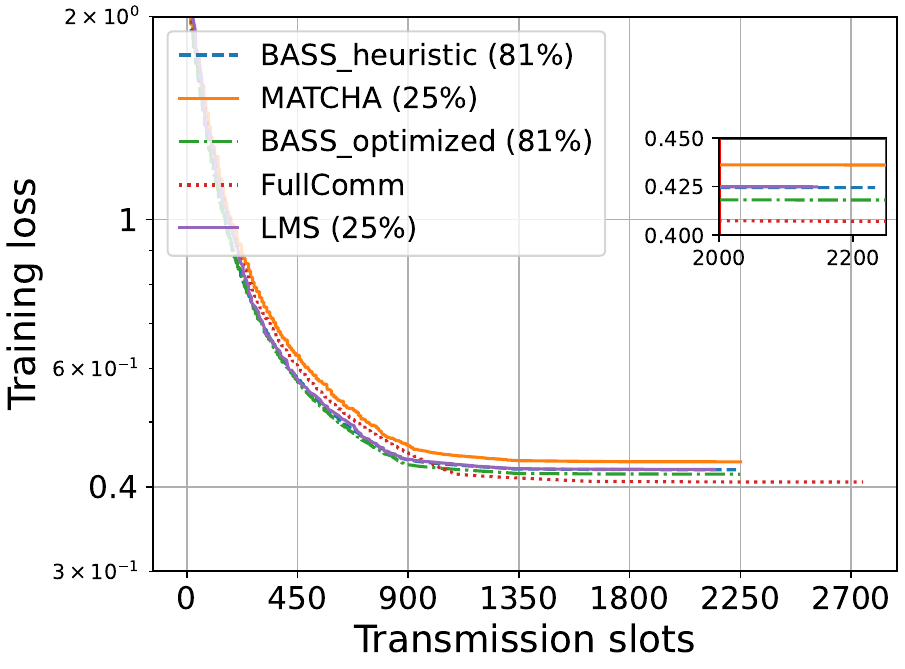}
	\subcaption{Training loss, ER}
	\vspace{0.5cm}
\end{subfigure}

\begin{subfigure}[c]{0.32\linewidth}
	\includegraphics[width=\columnwidth]{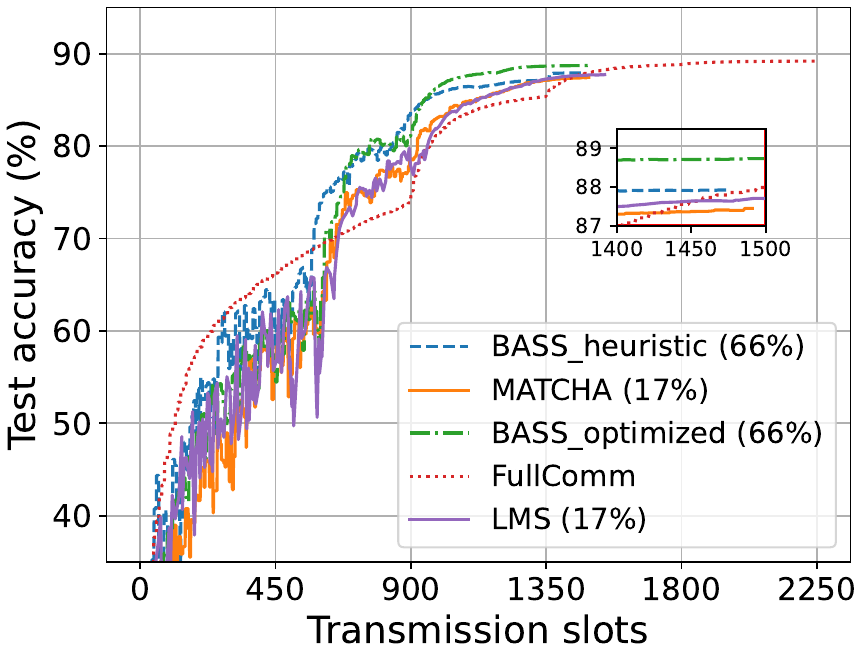}
	\subcaption{Test accuracy, geometric}
	\vspace{0.5cm}
\end{subfigure}
\hfill
\begin{subfigure}[c]{0.32\linewidth}
	\includegraphics[width=\linewidth]{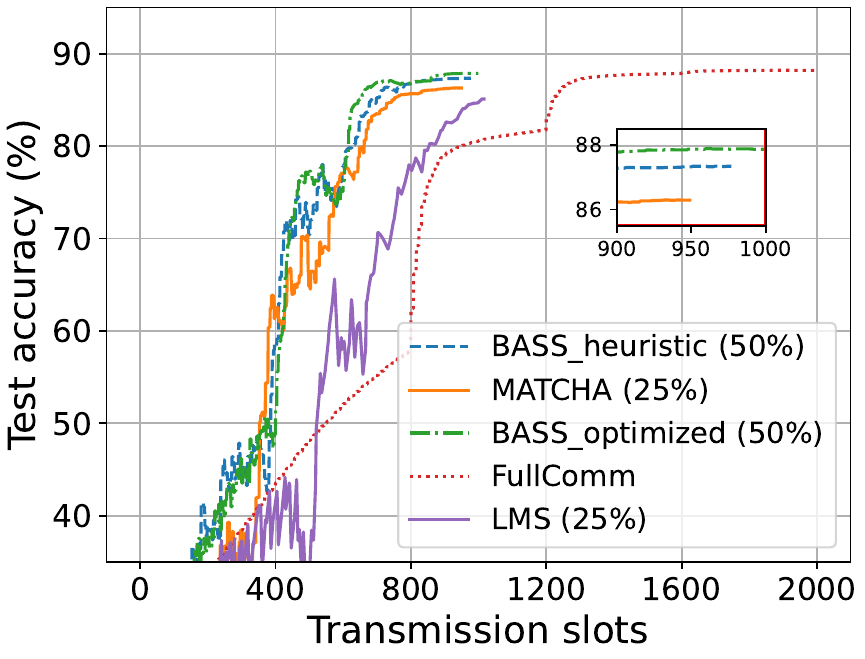}
	\subcaption{Test accuracy, two-stars}
	\vspace{0.5cm}
\end{subfigure}
\hfill
\begin{subfigure}[c]{0.32\linewidth}
	\includegraphics[width=\linewidth]{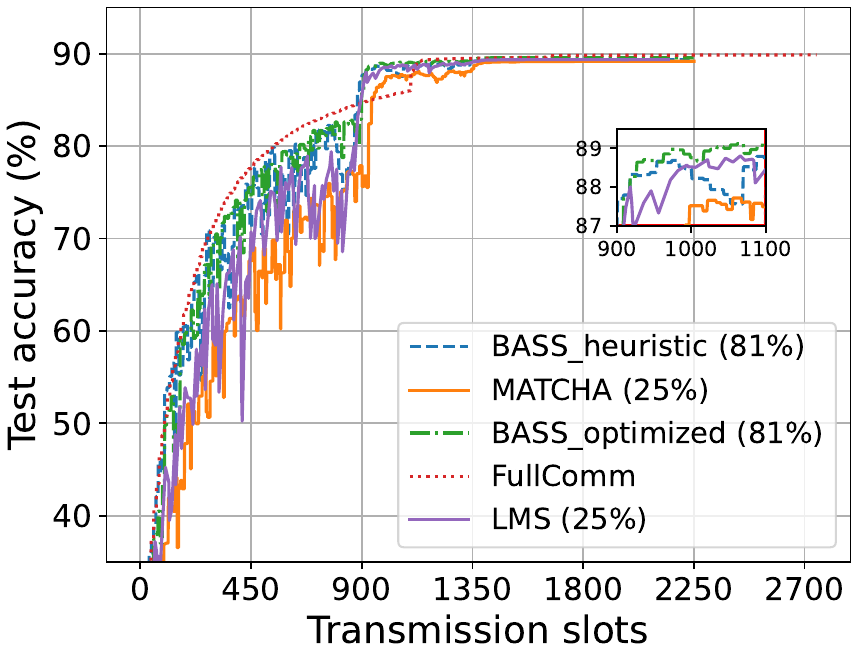}
	\subcaption{Test accuracy, ER}
	\vspace{0.5cm}
\end{subfigure}

\caption{Performance comparison between optimized and heuristic $\texttt{BASS}$, modified $\texttt{MATCHA}$, modified $\texttt{LMS}$, and full communication with different network typologies. .}
\label{exp_fig}
\end{figure*}

\section{Numerical Experiments}	
In this section, we evaluate the performance of our proposed broadcast-based subgraph sampling method for D-SGD with communication cost constraints. Simulations are performed by training a multi-layer perceptron (MLP) for an image classification task on MNIST datatset \cite{lecun2010mnist}. The model consists of a flattening layer that converts the $28\times28$-pixel input images into a $784$-dimensional vector. This is followed by a fully connected layer with $128$ units and ReLU activation. To prevent overfitting, we added a dropout layer with a dropout rate of $0.5$. The output layer is a fully connected layer with 10 units and softmax activation, corresponding to the $10$ classes in MNIST.  This dataset contains $60000$ images for training and $10000$ for testing.  The training dataset was divided into $2N$ shards (for a total of $N$ agents), and each agent's local dataset is given by $2$ randomly selected shards with no repetition. This ensures that the data is non-iid with a good level of heterogeneity.

We used \textit{SparseCategoricalCrossentropy} from Keras as
loss function and SGD optimizer. The initial learning rate used was
$0.05$, and it was reduced by $10\times$ after iterations $100, 150,
200$. The batch size is selected such that in every epoch, the
number of batches is equal to $5$, and the communication occurs
after each epoch. All experiments run for $250$ communication rounds
(iterations).
The performance of our proposed design is evaluated by showing the improvement of training loss and test accuracy with the number of transmission slots.
For performance comparison, we also implemented a modified version of $\texttt{MATCHA}$, a modified version of $\texttt{LMS}$, and the full communication case, where all nodes are activated in every round.  
For $\texttt{LMS}$, we created over $3000$ candidates using algorithm 3 of \cite{chiu2023laplacian}.
All experiments parameters were optimized for the full communication case.

\subsection{Performance Evaluation of $ \texttt{BASS}$}
\begin{figure}
    \centering
    \includegraphics[width=0.9\linewidth]{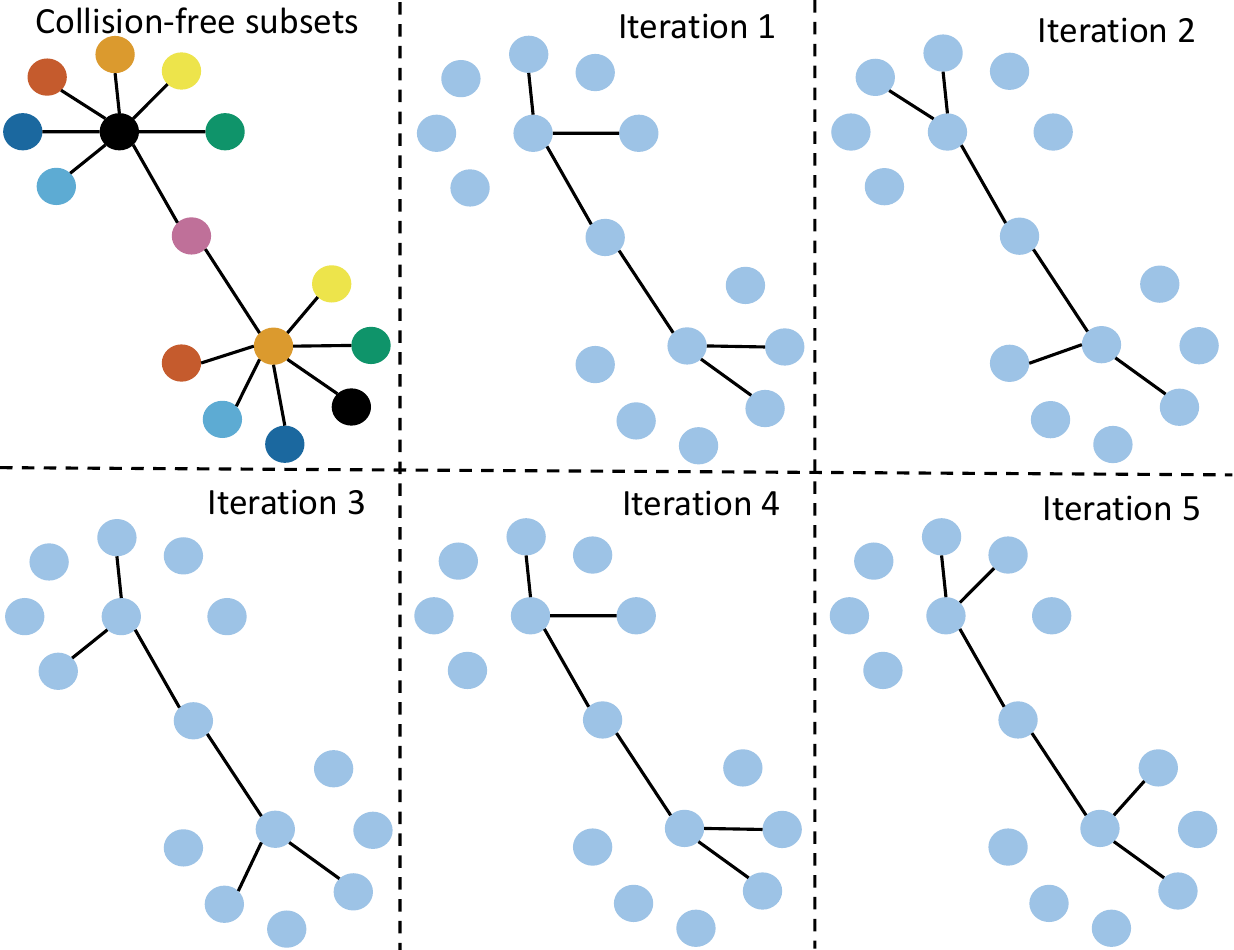}
    \caption{Collision-free subsets and subgraph sampling visualization for the topology in Fig. \ref{exp_fig}(b). The communication budget is $\mathcal{B}=4$, corresponding to the $50\%$ of the total number of collision-free subsets.}
    \label{subgraph_sampling}
\end{figure}

In Fig. \ref{exp_fig}, we compare the performance of optimized and heuristic $ \texttt{BASS}$, modified $\texttt{MATCHA}$ and $\texttt{LMS}$, and full communication for different topologies. 
Note that full communication is a special case of $ \texttt{BASS}$ if all subsets are scheduled in every round. We refer to this case as ``FullComm''.
The percentage next to $\texttt{MATCHA}$, $\texttt{LMS}$, and $ \texttt{BASS}$ represents the fraction of the total number of matchings and subsets that are activated on average per round, respectively. For the experiments, we test different activation percentages, and the one that gives the best performance is plotted in the figure. To better illustrate our proposed design, in Fig. \ref{subgraph_sampling} we show an example of subgraph sampling for the topology presented in Fig. \ref{exp_fig}(b).

We observe that optimized $ \texttt{BASS}$ clearly outperforms modified $\texttt{MATCHA}$ and modified $\texttt{LMS}$, since the former takes advantage of broadcast transmissions, allowing us to use less transmission slots to activate more links. We can also observe a gain in using optimized $ \texttt{BASS}$ over heuristic $ \texttt{BASS}$. 
Note that heuristic $\texttt{BASS}$ also outperforms modified $\texttt{MATCHA}$ and modified $\texttt{LMS}$.

\subsection{Impact of the Communication Budget}
\begin{figure}[t!]
\centering
\includegraphics[width=0.95\columnwidth]{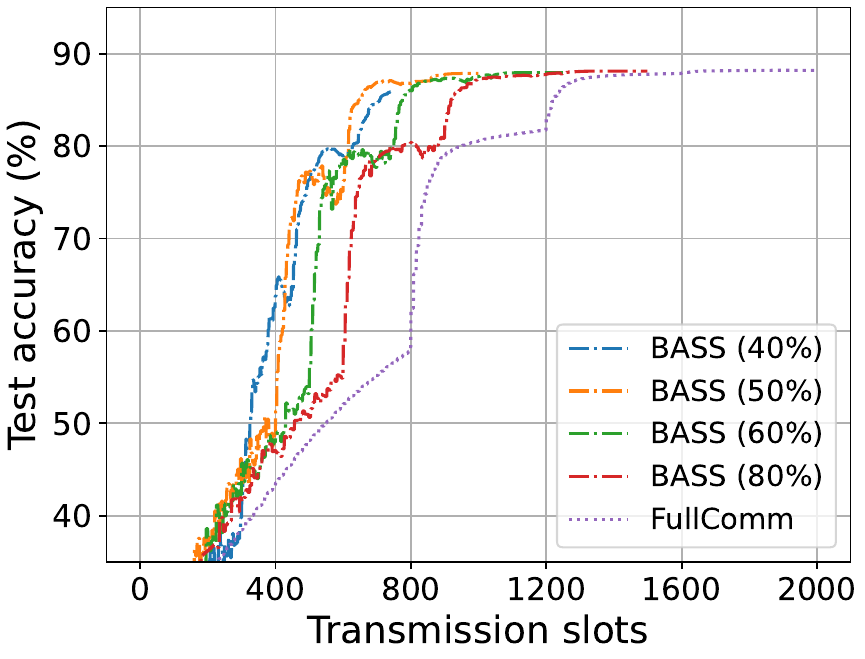}
\caption{Impact of the communication budget in the performance of $ \texttt{BASS}$.}
\label{comm_bud_comparison}
\end{figure}

The communication budget per iteration has an impact on the performance of $ \texttt{BASS}$, as shown in Fig. \ref{comm_bud_comparison}.
Here, we only consider optimized $ \texttt{BASS}$ (referred to as $ \texttt{BASS}$), and we use the graph in Fig. \ref{exp_fig}(b) as base topology. 
We can see that different communication budgets (the percentages in the figure) give very different results. On one hand, if the communication budget is low (small activation percentage), the number of activated links per iteration is also low, leading to poor information fusion/mixing and reducing performance. 
On the other hand, if the budget is very high, the communication improvement per transmission slot becomes marginal, leading to inefficient use of resources.

\subsection{Impact of Initial Graph Partition}
\begin{figure}[t!]
\centering
\includegraphics[width=0.95\columnwidth]{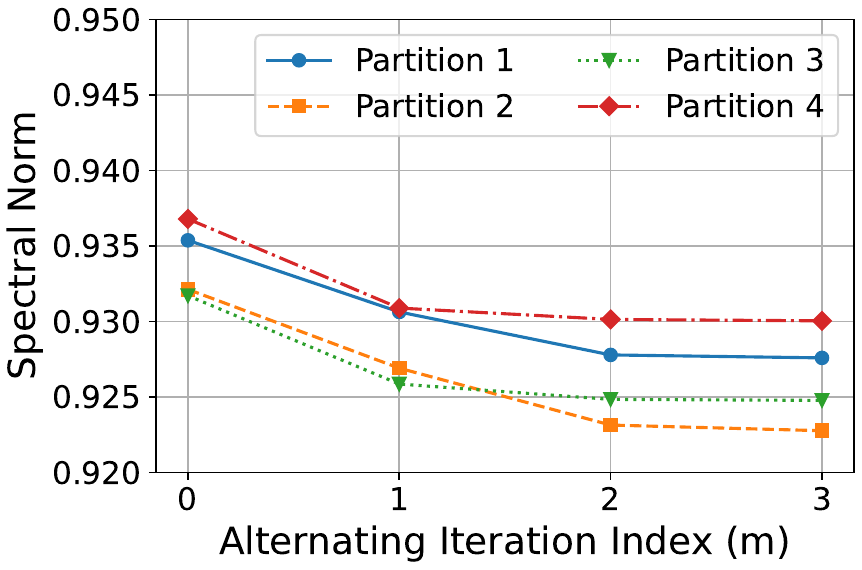}
\caption{Spectral norm evolution for different base topology partitions.}
\label{initial_partition_impact}
\end{figure}

With $\texttt{BASS}$, we first create collision-free subsets by partitioning the base topology. 
This partition result is not unique, and the final subgraph candidates obtained from the partitioned subsets will depend on the solution of the initial partitioning algorithm. We may improve the performance of $\texttt{BASS}$ by constructing multiple graph partitioning solutions and create all possible combinations of collision-free subsets. This will cause an increasing number of candidate subgraphs, and more optimization problems to solve to find the best communication pattern and mixing design.
In Fig. \ref{initial_partition_impact}, we observe the impact of considering different initial graph partitions on the spectral norm $\left\|\mathbb{E}[\matr{W}^{\top}(t)\matr{W}(t)]-\matr{J}\right\|_2$,  using the network topology in Fig. \ref{exp_fig}(a) as base topology. 

\subsection{Impact of Network Density}
\begin{figure}[t!]	
\begin{subfigure}[c]{0.49\columnwidth}
	\includegraphics[width=\columnwidth]{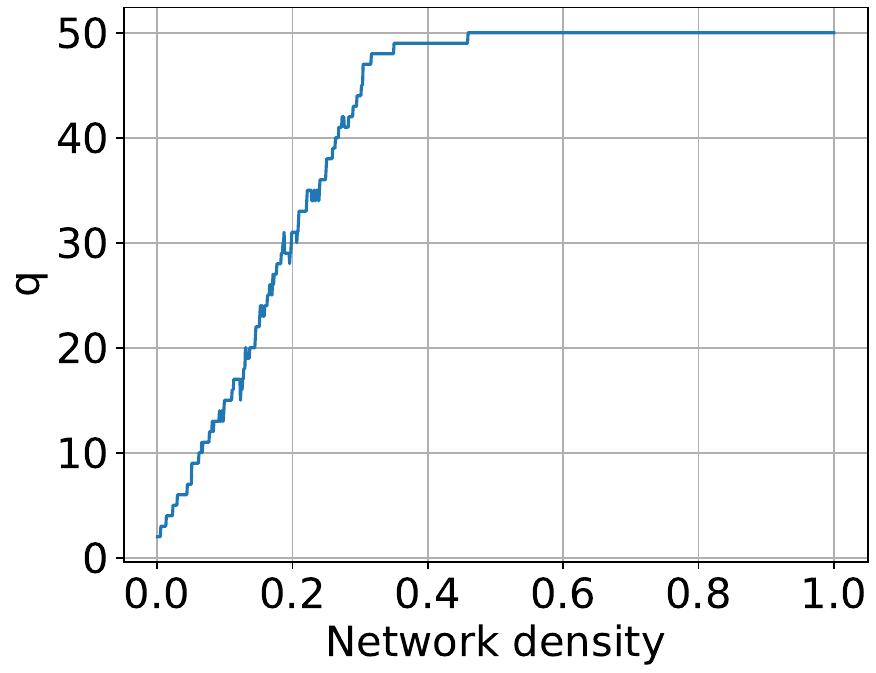}
	\subcaption{Network Density vs. $q$}
\end{subfigure}	  
\hfill
\begin{subfigure}[c]{0.5\columnwidth}
	\includegraphics[width=\columnwidth]{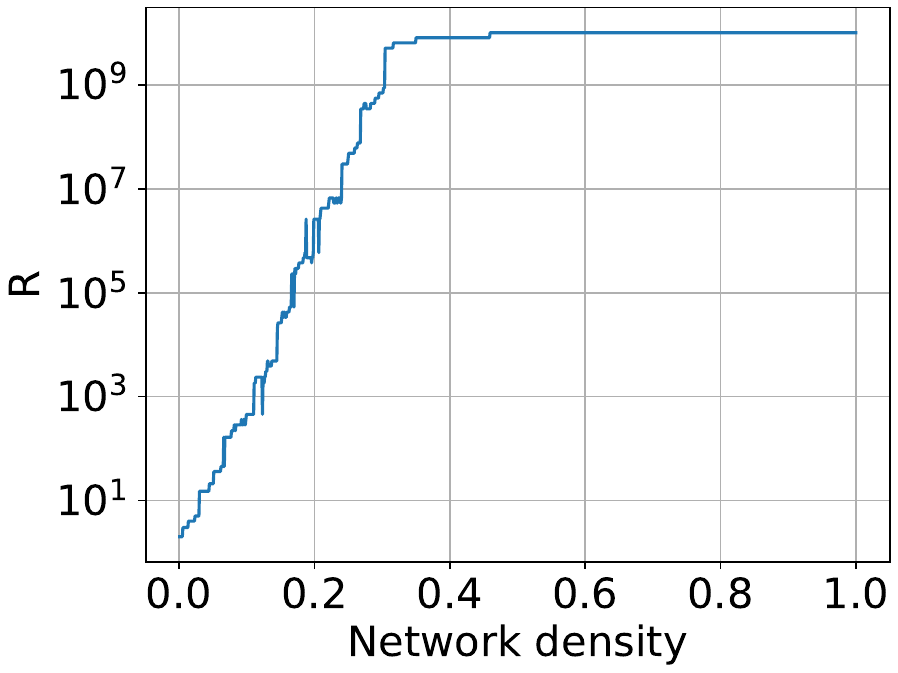}
	\subcaption{Network Density vs. $R$}
\end{subfigure}	
\caption{Impact of network density on the number of collision-free subsets $q$ and the number of subgraph candidates $R$ for a fixed budget $\mathcal{B}=10\%$. The network has $50$ nodes.}
\label{impact_of_density}
\end{figure}

For a given partition of the base graph with $q$ collision-free subsets, and a communication budget $\mathcal{B}$, we can find all possible subgraphs whose activation consumes exactly $\mathcal{B}$ transmission slots by taking $R=$ $q\choose \mathcal{B}$. This shows the relationship between the number of subgraphs $R$, the communication budget $\mathcal{B}$, and the number of collision-free subsets $q$. By construction, $\mathcal{B}$ is a hyperparameter that can be tuned to improve performance, but $q$ is network dependent. For instance, the more densely connected the network is, the larger the number of collision-free subsets, increasing the number of candidates for a fixed budget. The network density is defined as:
\begin{equation}
    \text{Network Density} = \frac{\mathbbm{1}^\top\matr{A}\mathbbm{1}}{N(N-1)},
\end{equation}
In Fig. \ref{impact_of_density}(a) below we can see that the number of collision-free subsets is equal to the number of nodes when the network density is $0.5$ or above. Similarly, we can see in Fig. \ref{impact_of_density}(b) that the number of subgraph candidates becomes very large for denser graphs. This indicates that our algorithm is more efficient for sparser network topologies. However, in denser topologies, we can reduce the number of candidates by only considering a fraction of the total number of subgraph candidates, as long as the union of the selected subgraphs gives a connected topology.

\subsection{Impact of Link Failures}
\begin{figure}
    \centering
    \includegraphics[width=0.95\linewidth]{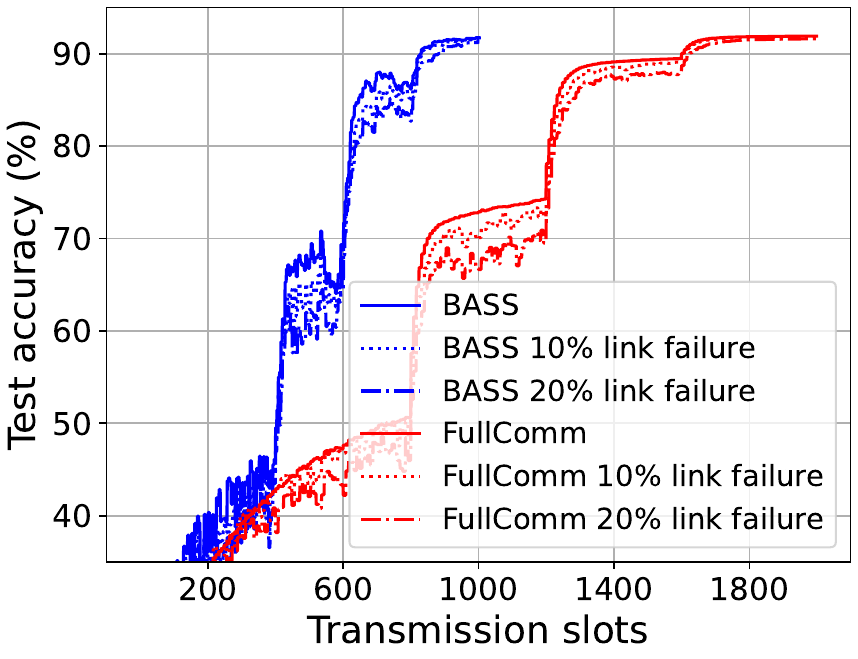}
    \caption{Comparison of perfect and imperfect communication for the topology in Fig. 3(b). The communication budget for $\texttt{BASS}$ is $50\%$. With imperfect communication, each link has the same failure probability, and the link failures are bi-directional and mutually independent.}
    \label{imperfect_vs_perfect_comm}
\end{figure}

D-SGD in the presence of link failures introduces an extra source of randomness that reduces the performance as compared to the case of perfect communication. There are several techniques to deal with the missing information, such as the biased compensation method in \cite{fagnani2009average}, where each agent adds the weight of the failed links to its self-weight. With this technique, we ensure that the sum of each row of the mixing matrix remains equal to one, and if the mixing matrix is symmetric (as considered in this work), the sum of each column is also equal to one.
As we can see in Fig. \ref{imperfect_vs_perfect_comm}, for a uniform link failure probability of $10\%$, or even $20\%$, the performance reduction is not significant, and the same test accuracy is achieved, indicating the robustness of our method in the presence of link failures.

\subsection{Additional Experiments}

\subsubsection{CIFAR-10 dataset}
\begin{figure}[t!]
\centering
\includegraphics[width=0.95\columnwidth]{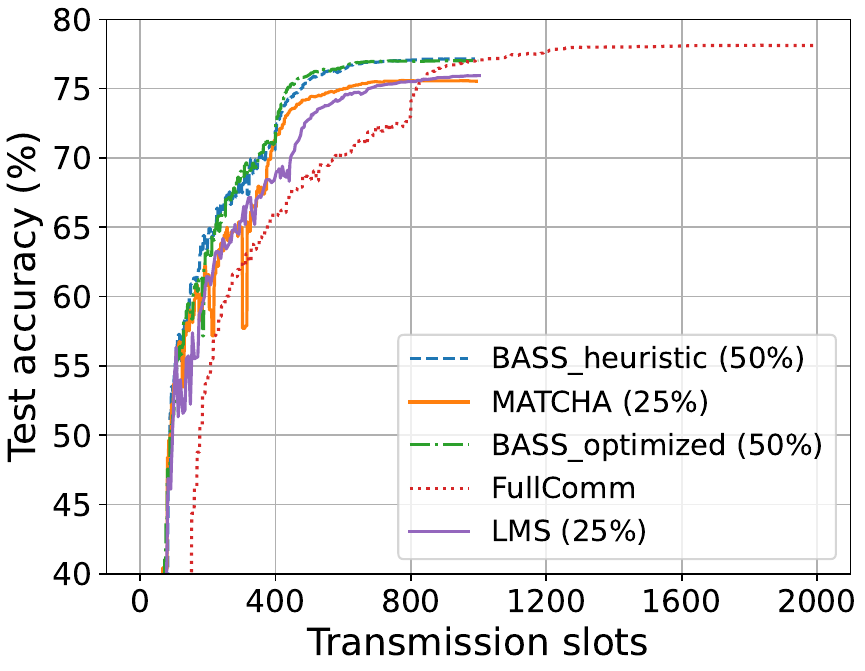}
\caption{Performance comparison of different methods using CIFAR-10 dataset.}
\label{cnn_performance}
\end{figure}

In addition to MNIST, we also validate the performance of our proposed method on CIFAR-10 dataset \cite{krizhevsky2009learning}. In Fig. \ref{cnn_performance}, we show the performance comparison of all studied methods using the graph in Fig. \ref{exp_fig}(b) as base topology. Similar to the results obtained with MNIST, We observe obvious performance gains of our proposed methods over $\texttt{MATCHA}$ and $\texttt{LMS}$ on CIFAR-10 dataset.

The trained model consists of a convolutional neural network (CNN) with residual blocks for image classification on the CIFAR-10 dataset. The model has an input layer accepting images of size $32\times32\times3$. The first convolutional layer applies $32$ filters of size $3\times3$ with ReLU activation.
We add residual blocks where each block consists of two convolutional layers, each with filters of size $3\times3$ and ReLU activation, followed by batch normalization. A shortcut connection adds the input of the block to the output of the second convolutional layer, implementing the residual connection.
After the initial convolutional layer, we added a residual block with $32$ filters, followed by a max pooling layer with a pool size of $2\times2$ and a dropout layer with a dropout rate of $0.25$.
This structure is repeated with $64$ filters: a convolutional layer, a residual block, a max pooling layer with a pool size of $2\times2$, and a dropout layer with a dropout rate of $0.25$.
The feature maps are then flattened and passed through a fully connected layer with $512$ units and ReLU activation, followed by a dropout layer with a dropout rate of $0.5$. The output layer is a fully connected layer with $10$ units and softmax activation, corresponding to the $10$ classes in CIFAR-10.
We used the Adam optimizer with an initial learning rate of $0.001$ and trained the model for $250$ epochs with a batch size of $128$.

\vspace{-0.4cm}
\subsubsection{Large networks}
\begin{figure}[t!]	
\begin{subfigure}[c]{0.49\columnwidth}
	\includegraphics[width=\columnwidth]{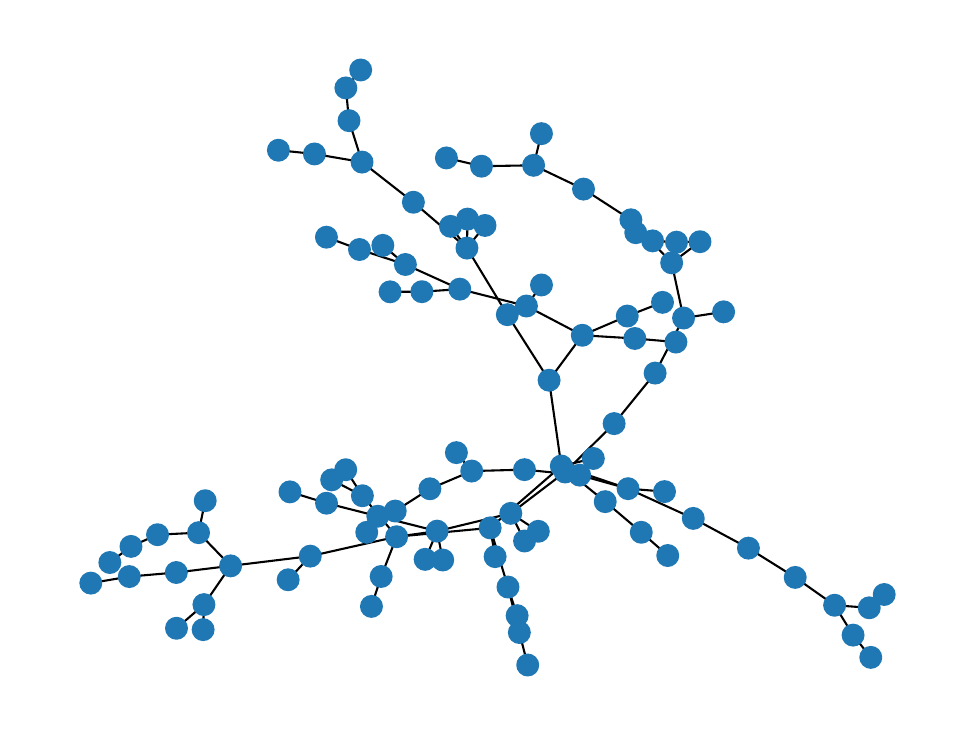}
	\subcaption{Forest graph}
\end{subfigure}	  
\hfill
\begin{subfigure}[c]{0.5\columnwidth}
	\includegraphics[width=\columnwidth]{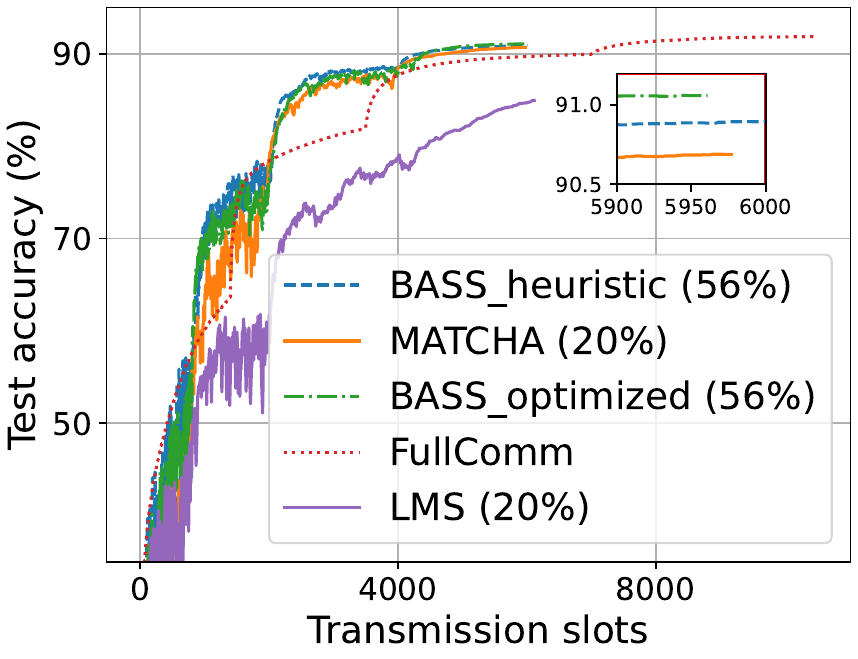}
	\subcaption{Test accuracy }
\end{subfigure}	
\caption{Performance comparison of different methods on a 100-node graph. (a) shows the graph topology, (b) shows the test accuracy obtained with this graph.}
\label{forest_comp}
\end{figure}

In Fig. \ref{forest_comp}, we show the performance comparison of optimized and heuristic $\texttt{BASS}$, FullComm, $\texttt{MATCHA}$, and $\texttt{LMS}$ for a network of $100$ nodes. 
Similar to previous cases, optimized $\texttt{BASS}$ outperforms all other approaches. This shows the scalability of our proposed design.

\section{Conclusions and Future Directions}
This work focuses on broadcast-based multi-access communication design for accelerating the convergence of decentralized learning over networked agents, considering the actual communication delay per iteration. Our proposed framework, $\texttt{BASS}$, generates a family of mixing matrix candidates associated with a set of sparser subgraphs and their sampling probabilities. In each iteration of the D-SGD algorithm, one mixing matrix is randomly sampled. The corresponding subgraph indicates which set of nodes can broadcast their current models to their neighbors. Unsurprisingly, $\texttt{BASS}$ outperforms existing link-based scheduling policies and plain D-SGD by a considerable margin, achieving faster convergence with fewer transmission slots.

Our findings underscore two key points: 1) Identifying and sampling subgraphs of the base topology can be exploited to customize communication dynamics and improve the error-vs-runtime tradeoff, and 2) Leveraging broadcast transmission and spatial reuse of resources can accelerate the convergence of D-SGD over wireless networks.
A potential direction for future exploration involves jointly considering the impact of network connectivity and data heterogeneity in node scheduling design. 
Another potential extension is to optimize the communication budget  allocation over time, using an adaptive or event-triggered design.

\section*{Appendix}
\vspace{-0.1cm}
\subsection{Poof for the Convexity of Problem \eqref{joint_optimization} Under a Given Set of Sampling Probabilities}
\label{sec:proof}
The objective function of problem (\ref{joint_optimization}) can be written as:
\begin{equation}
\left\|\mathbb{E}[\matr{W}^{\top}(t)\matr{W}(t)]-\matr{J}\right\|_2 = \lambda_{\max}\left(\sum_{r=1}^{R}p_r\matr{W}_r^{\top}\matr{W}_r-\matr{J}\right) \nonumber\\
\end{equation}
\begin{align}
= & \sup_{||\vect{x}||_2=1} \vect{x}^{\top}\left(\sum_{r=1}^{R}p_r\matr{W}_r^{\top}\matr{W}_r-\matr{J}\right)\vect{x} \nonumber\\
= & \sup_{||\vect{x}||_2=1} \sum_{r=1}^{R}p_r \vect{x}^{\top}\matr{W}_r^{\top}\matr{W}_r \vect{x} - \vect{x}^{\top}\matr{J}\vect{x} \nonumber\\
= & \sup_{||\vect{x}||_2=1} \sum_{r=1}^{R}p_r \text{Tr}\left(\matr{W}_r \vect{x} \vect{x}^{\top}\matr{W}_r^{\top}\right) - \vect{x}^{\top}\matr{J}\vect{x}\nonumber \\
= &\sup_{||\vect{x}||_2=1} \sum_{r=1}^{R} p_r \sum_{j=1}^{N} (\vect{w}_j^r)^{\top} \vect{x} \vect{x}^{\top} \vect{w}_j^r - \vect{x}^{\top} \matr{J} \vect{x}
\label{quadratic_form}
\end{align}
where $\vect{w}_j^r$ represents the $j$-th column of the matrix $\matr{W}_r$. 
The function $f(\vect{w}) = \vect{w}^{\top}\vect{x}\vect{x}^{\top}\vect{w}$ is a quadratic form with kernel $\vect{x}\vect{x}^{\top}$, which is positive semidefinite; therefore, $f(\vect{w})$ is convex for any fixed $\vect{x}$. 
Applying the previous reasoning to (\ref{quadratic_form}) and the pointwise supremum principle \cite[Section 3.2.3]{boyd2004convex}, we conclude that the objective function is jointly convex with respect to all candidates $\{\matr{W}_r\}_{r\in[R]}$.

\bibliographystyle{IEEEtran}

\end{document}